\documentclass{eptcs}
\providecommand{\event}{PrePost 2017}

\usepackage{xspace}
\usepackage[utf8]{inputenc}
\usepackage{xargs}

\usepackage[utf8]{inputenc}
\usepackage[usenames,dvipsnames,svgnames,table]{xcolor} % must be loaded before tikz
\usepackage{tikz}

\usepackage[nomargin,multiuser,inline,draft]{fixme} 
\fxusetheme{color}
\usepackage{xargs}

\usepackage{wrapfig}

\FXRegisterAuthor{ic}{ian}{\color{blue}   {\underline{Ian}}}
\FXRegisterAuthor{af}{AF}{\color{red}   {\underline{AF}}}
\FXRegisterAuthor{cm}{CM}{\color{ForestGreen}   {\underline{CM}}}
\FXRegisterAuthor{eM}{aeM}{\color{orange} {\underline{eM}}}

\usepackage{mathtools}
\usepackage{amsmath}
\usepackage{amsthm}

\newtheorem*{example}{Example}

\usepackage{amssymb}
\usepackage{caption}                % Needed by subcaption
\usepackage{cmll} % \bigwith symbol
\usepackage{colonequals}            % More symbols (colon equalities etc.)
\usepackage{environ}                % Provides NewEnviron w/ eager expansion
\usepackage{etex,etoolbox}          % Needed by theorem-appendix.tex
\usepackage{graphicx}               % For \scalebox etc.

\usepackage[protrusion=true,expansion=true]{microtype} % Better typography
\usepackage{multirow}               % Table multirow/multicolumn support
\usepackage{hyperref}
\usepackage[capitalise]{cleveref}
\usepackage{nicefrac}
\usepackage{proof}                  % Logic proofs
\usepackage{bussproofs}
\EnableBpAbbreviations

\usepackage{rotating}               % For \sidewaysfigure etc.
\usepackage{subcaption}             % Support for subfigures
\usepackage{thmtools, thm-restate, thm-autoref}  % Advanced theorem handling
\usepackage{xifthen}                % for conditional commands
\usepackage{appendix}               % for correct numeration of statements

\usepackage{stmaryrd}
 \usepackage{pdfsync}

\usepackage[usenames,dvipsnames,svgnames,table]{xcolor} % must be loaded first
\usepackage{tikz}                   % Graphs
\usetikzlibrary{shadows,arrows,shapes,automata,positioning,decorations.markings}

\usepackage[nomargin,inline]{fixme} % Simplified management of FIXME's
\fxusetheme{color}

\usepackage{mathtools} % \xRightArrow and stuff - load late to avoid conflicts
\usepackage{amsbsy} % for bold math symbols

\DeclareMathSizes{10}{10}{6}{6}

\def\colorPtp{\color{Blue}}

\def\colorOp{\color{ForestGreen}}
\def\colorNode{\color{cyan}}
\def\colorFun{\color{red}}

\def\colorE{\color{orange}}

\def\colorMsg{\color{BrickRed}}

\newcommand{\msg}[1][m]{\mathsf{\colorMsg{#1}}}

\newcommand{\ptp}[1][A]{\ensuremath{\mathsf{\colorPtp{\MakeUppercase{#1}}}}}

\newcommand{\p}{\ptp}
\newcommand{\q}{{\ptp[B]}}

\newcommandx{\common}[3][1=\ptp,2={\aH},3={\aH'},usedefault=@]{f_{#1}}
\newcommandx{\opair}[2][1={\ae},2={\ae'},usedefault=@]{\conf{{#1},{#2}}}
\newcommandx{\hopair}[2][1={\aE},2={\aE'},usedefault=@]{\llparenthesis\, {#1},{#2}\, \rrparenthesis}
\newcommandx{\wb}[2][1={\aG},2={\aG'},usedefault=@]{wb({#1}, {#2})}
\newcommandx{\ws}[2][1={\aseq},2={\aseq'},usedefault=@]{ws({#1}, {#2})}
\newcommandx{\widx}[2][1={\aW},2={i},usedefault=@]{{#1}[{#2}]}
\newcommandx{\outop}[2][1=\gname,2={}]{{\colorOp{!}}^{{#1}{#2}}}
\newcommandx{\inop}[2][1=\gname,2={}]{{\colorOp{?}}^{{#1}{#2}}}
\newcommandx{\aout}[5][1={\p},2={\q},3=\gname,4=\msg,5={},usedefault=@]{
  \achanl[{#1}][{#2}]{#5} \outop[{#3}] {#4}{#5}
}
\newcommandx{\ain}[5][1={\p},2={\q},3=\gname,4=\msg,5={},usedefault=@]{
  \achanl[{#1}][{#2}]{#5} \inop[{#3}] {#4}{#5}
}
\newcommandx{\adep}[1][1={}]{
  \conf{ \aout[@][@][@][@][{#1}], \ain[@][@][@][@][{#1}]}
}

\newcommandx{\hproj}[2][1=\aH, 2=\ptp, usedefault=@]{
  \ifempty{#1}{}{{#1}}\ifempty{#2}{}{{^{\scriptscriptstyle @{#2}}}}
}
\newcommandx{\eproj}[2][1=\aE,2=\ptp, usedefault=@]{
  {{#1}}\ifempty{#2}{}{{^{\scriptscriptstyle @{#2}}}}
}

\newcommand{\aM}{M}
\newcommandx{\cm}[2][1=\ptp, 2=\aM]{{#2}_{#1}}
\newcommandx{\achan}[2][1=A,2=B,usedefault=@]{\ptp[#1]\!\cdot\!\ptp[#2]}
\newcommandx{\achanl}[2][1=\q,2=\q,usedefault=@]{#1\cdot#2}

\newcommand{\oact}{\outop[]}
\newcommand{\iact}{\inop[]}
\newcommand{\tset}{\to}

\newcommandx{\acfsmout}[3][1=A,2=B,3=m,usedefault=@]{\achan[{\ptp[{#1}]}][{\ptp[{#2}]}] \oact {\msg[{#3}]}}
\newcommandx{\acfsmin}[3][1=A,2=B,3=m,usedefault=@]{\achan[{\ptp[{#1}]}][{\ptp[{#2}]}] \iact {\msg[{#3}]}}
\newcommandx{\fsaout}[2][1={\p},2={},usedefault=@]{
  \ptp[{#1}] \ \outop[]\ \msg[{#2}]
}
\newcommandx{\fsain}[2][1={\p},2={},usedefault=@]{
  \ptp[{#1}] \ \inop[]\ \msg[{#2}]
}

\newcommand{\aG}{\mathsf{G}}
\newcommand{\gsink}{\circledcirc}
\newcommand{\gsource}{\circ}
\newcommand{\gvertex}{\bullet}
\newcommand{\gseqop}{{\colorOp ;}}
\newcommand{\gparop}{{\colorOp \,|\,}}
\newcommand{\gchoop}{{\colorOp +}}
\newcommand{\grecop}{{\colorOp *}}
\newcommand{\grecopp}{{\colorOp{@}}}
\newcommand{\gname}[1][i]{{\colorNode{\scriptstyle\textsf{#1}}}}
\newcommand{\nmerge}[1]{\mbox{\colorNode -}{#1}}

\newcommandx{\gnode}[2][1=\gname,2=\gint,usedefault=@]{
  {\ifempty{#1}{}{\colorNode{#1.}}} {#2}
}

\newcommandx{\gint}[4][1=\gname,2=\ptp,3=\msg,4={\ptp[B]},usedefault=@]{
  {#2} {\colorOp \xrightarrow{#1}} {#4} \colon {\msg[{#3}]}
}
\newcommandx{\gout}[4][1=\gname,2=\ptp,3=\msg,4={\ptp[C]},usedefault=@]{
  \achan[{#2}][{#4}] {\colorOp {\colorOp{!}}} {\msg[{#3}]}
}
\newcommandx{\gin}[4][1=\gname,2=\ptp,3=\msg,4={\ptp[C]},usedefault=@]{
  \achan[{#2}][{#4}] {\colorOp {\colorOp{?}}} {\msg[{#3}]}
}
\newcommandx{\gseq}[3][1=\gname,2={\aG},3={\aG'},usedefault=@]{
  \gnode[{#1}][{#2} \gseqop {#3}]
}
\newcommandx{\gpar}[3][1=\gname,2={\aG},3={\aG'},usedefault=@]{
  \gnode[{#1}][\ifempty{#1}{{#2} \gparop {#3}}{({#2} \gparop {#3})}]
}
\newcommandx{\gcho}[3][1=\gname,2={\aG},3={\aG'},usedefault=@]{
  \gnode[{#1}][\ifempty{#1}{{#2} \gchoop {#3}}{\big({#2} \gchoop {#3}\big)}]
}
\newcommandx{\grec}[3][1=\gname,2={\aG},3={\p},usedefault=@]{
  \gnode[{#1}][\ifempty{#1}{\grecop {#2} \grecopp {#3}}{\big(\grecop {#2} \grecopp {#3}\big)}]
}

\newcommandx{\gsem}[2][1={\aG},2={},usedefault=@]{[\![ {#1} ]\!]_{#2}}
\newcommandx{\rbot}{\text{undef}}
\newcommandx{\rtrs}[1][1={\aH},usedefault=@]{{#1}^{\star}}
\newcommandx{\gord}[1][1={\aG},usedefault=@]{\leq_{#1}}
\newcommandx{\gordeq}[1][1={\aG},usedefault=@]{\leq_{#1}}

\newcommandx{\rlang}{\mathcal{L}}

\newcommandx{\aW}{w}

\newcommand{\gfun}[1]{\ensuremath{\mathsf{\colorFun #1}}}

\newcommandx{\rseq}[2][1=\aG,2={\aG'},usedefault=@]{\gfun{seq}({#1},{#2})}
\newcommandx{\rpar}[2][1=\aG,2={\aG'},usedefault=@]{\gfun{par}({#1},{#2})}

\newcommandx{\gproj}[2][1=\aG,2=\ptp]{{#1}\downarrow_{#2}}
\newcommandx{\cinit}[1][1={\aQzero},usedefault=@]{{#1}}
\newcommandx{\cfinal}[1][1={q_e},usedefault=@]{{#1}}

\newcommandx{\geproj}[4][1=\aG,2=\ptp,3=\cinit,4=\cfinal,usedefault=@]{
  {#1}\downarrow_{#2}^{{#3},{#4}}
}

\newcommand*{\StrikeThruDistance}{0.15cm}%
\tikzset{strike thru arrow/.style={
    decoration={markings, mark=at position 0.5 with {
        \draw [blue, thick,-] 
            ++ (-\StrikeThruDistance,-\StrikeThruDistance) 
            -- ( \StrikeThruDistance, \StrikeThruDistance);}
    },
    postaction={decorate},
}}

\newcommandx{\ich}[1][1={\aG},usedefault=@]{{#1}^{\oplus}}
\newcommandx{\ichedges}[2][1={\aG},2={\gname},usedefault=@]{{#1}^{\oplus}({#2})}
\newcommandx{\parts}[1]{2^{#1}}
\newcommandx{\actch}{c}
\newcommandx{\soundactch}[2][1={\aG},2={\actch},usedefault=@]{{#1} \,\circledR\, {#2}}
\newcommandx{\rOnActch}[2][1={\aG},2={\actch},usedefault=@]{{#1} \setminus {#2}}
\newcommandx{\rOnActchClean}[2][1={\aG},2={\actch},usedefault=@]{{#1} \circledR {#2}}
\newcommandx{\rAllEvents}[1][1={\aG},usedefault=@]{\mathit{dom}(#1)}

\newcommand{\aH}{H}

\newcommandx{\hgvertex}[2][1=\al,2=\gname,usedefault=@]{{#1}_{\textcolor{red}{[{#2}]}}}
\newcommand{\aE}{{\tilde \ae}}

\renewcommand{\ae}[1][e]{{\colorE{#1}}}
\newcommand{\al}[1][l]{{\colorE{#1}}}
\newcommandx{\hyedge}[1]{\{#1\}}

\newcommandx{\rdiv}[2][1=\gcho,2=\ptp,usedefault=@]{
  \gfun{div}_{#2}(#1)
}

\newcommandx{\rrdiv}[5][1={\aG},2={\aG'},3={\aE},4={\aE'},5=\ptp,usedefault=@]{
  \gfun{div}^{#3,#4}_{#5}(#1,#2)
}
\newcommandx{\pdiv}[3][1={\apom_1},2={\apom_2},3={\apom},usedefault=@]{
  \gfun{div}_{#3}(#1,#2)
}
\newcommandx{\pfork}[3][1={\apom_1},2={\apom_2},3={\apom},usedefault=@]{
  \gfun{fork}_{#3}(#1,#2)
}

\newcommand{\gatedistancein}{3pt}
\newcommand{\gatedistanceinand}{2pt}

\usetikzlibrary{
  arrows,shapes,automata,backgrounds,petri,positioning,fit,calc,
  decorations.markings,shadows,fadings,patterns,
  decorations.pathreplacing,chains
}

\tikzset{
  src/.style={draw,circle,fill=white,
    minimum size=2mm,
    inner sep=0pt
  },
  sink/.style={draw,circle,double,fill=white,
    minimum size=1.5mm,
    inner sep=0pt
  },
  node/.style={draw,circle,fill=black,
    minimum size=2mm,
    inner sep=0pt
  },
  source/.style={draw,circle,fill=white,
    minimum size=4mm,
    inner sep=0pt
  },
  sink/.style={draw,circle,double,fill=white,
    minimum size=3mm,
    inner sep=0pt
  },
  block/.style = {rectangle, draw=gray, align=center, fill=orange!25, rounded corners=0.1cm,
    minimum size=5mm, inner sep=2pt},
  prenode/.style = {minimum size=9pt,inner sep=2pt, font=\Large},
  bblock/.style = {rectangle, draw=blue!50, opacity=.5, line width=1pt, align=center, fill=white, rounded corners=0.1cm,
    minimum size=7mm, inner sep=2pt},
  prenode/.style = {minimum size=9pt,inner sep=2pt, font=\Large},
  agate/.style={draw, rectangle,
    minimum size=3mm,
    inner sep=0pt,
    fill=orange!25,
    postaction={path picture={% 
        \draw[red]
        ([yshift=\gatedistanceinand]path picture bounding box.south) --
        ([yshift=-\gatedistanceinand]path picture bounding box.north) ;}}},
  ogate/.style = {
    diamond, draw, fill=orange!25,
    minimum size=4mm,
    inner sep=0pt,
    postaction={path picture={% 
        \draw[red]
        ([yshift=\gatedistancein]path picture bounding box.south) -- ([yshift=-\gatedistancein]path picture bounding box.north)
        ([xshift=-\gatedistancein]path picture bounding box.east) -- ([xshift=\gatedistancein]path picture bounding box.west)
        ;}}},
  altogate/.style = {
    diamond, draw,
    minimum size=4mm,
    inner sep=0pt,
    postaction={path picture={% 
        \draw
        ([yshift=\gatedistancein]path picture bounding box.south) -- ([yshift=-\gatedistancein]path picture bounding box.north)
        ([xshift=-\gatedistancein]path picture bounding box.east) -- ([xshift=\gatedistancein]path picture bounding box.west)
        ;}}},
  altgate/.style={draw, rectangle,
    minimum size=3mm,
    inner sep=0pt,
    postaction={path picture={% 
        \draw
        ([yshift=\gatedistanceinand]path picture bounding box.south) --
        ([yshift=-\gatedistanceinand]path picture bounding box.north) ;}}},
  anygate/.style = {circle, draw, fill=white,
    minimum size=4mm,
    inner sep=0pt,
    postaction={path picture={% 
        \draw[black]
        ([xshift=-\gatedistancein,yshift=\gatedistancein]path picture bounding box.south east) --
        ([xshift=\gatedistancein,yshift=-\gatedistancein]path picture bounding box.north west)
        ([xshift=-\gatedistancein,yshift=-\gatedistancein]path picture bounding box.north east) --
        ([xshift=\gatedistancein,yshift=\gatedistancein]path picture bounding box.south west)
        ;}}
  },
  smallglobal/.style={
        node distance=1cm and 0.8cm, semithick, scale=0.8, every node/.style={transform shape}
  },
  elli/.style = {draw,densely dotted,-},
  line/.style = {draw,->, rounded corners=0.07cm,>=latex},
  nline/.style = {draw,semithick, ->},
  pline/.style = {draw,->,>=latex},
  node distance=1cm and 0.7cm,
  baseline=(current  bounding  box.center),
  mycfsm/.style={
        font=\footnotesize,
        initial where=above,
        ->,>=stealth,auto, node distance=1cm and 1cm,
        scale=1, every node/.style={transform shape},
        baseline=(current  bounding  box.center)
  },
  machinecloud/.style={
    cloud, cloud puffs=10, cloud ignores aspect, minimum height=.1cm, minimum width=2cm, draw
  }
}

\newcommand{\gunlessop}{\mbox{\colorOp\tiny\tt unless}}

\newcommandx{\gtry}[5][1=\gname,2={\aG_1 \gchoop \cdots \gchoop \aG_n},3=\gin,4=\gout,5={j},usedefault=@]{
  \def\foo{\gtryop\ {#2} \ \gcatchop\ {#3} {\colorOp \Rightarrow} {#4} {\colorOp \bullet} {\gname[{#5}]}}
  \gnode[{#1}][{\ifempty{#1} {\foo } { \big(\foo \big) }}]
}

\newcommandx{\gtrycatch}[4][1=\gname,2={\aG},3=\gin,4={\aG'},usedefault=@]{
  \def\foo{\gtryop\ {#2} \ \gcatchop\ {#3} \gdoop\ {#4}}
  \gnode[{#1}][{\ifempty{#1} {\foo} {\big( \foo \big) }}]
}

\def\colorGuard{\color{cyan}}
\newcommand{\aguard}{{\colorGuard \phi}}
\newcommandx{\agG}[2][1={\aG},2=\aguard]{{#1} \ifempty{#2}{}{\ \gunlessop\ {#2}}}

\newcommandx{\grcho}[4][1=\gname,2={\agG},3={\agG[\aG'][\aguard']},4={\cdots},usedefault=@]{
  \def\foo{{#2} {\ \ifempty{#4}{\gchoop}{\gchoop \ {#4}\  \gchoop}\ } {#3}}
  \gnode[{#1}][\ifempty{#1}{\foo}{\big( \foo \big)}]
}

\newcommand{\aconfigfn}{\chi}
\newcommand{\aconfig}{\ell}

\newcommand{\lstates}{\statemap}
\newcommandx{\sysconfig}[3][1=\lstates,2=\aconfigfn,3={},usedefault=@]{
  \conf{ {#1},{#2} \ifempty{#3}{}{, #3} }
}
\newcommand{\sysctxfn}[1][]{\gamma_{#1}}
\newcommandx{\sysctx}[2][1=\aQ,2={},usedefault=@]{({#1},\sysctxfn[{#2}])}

\newcommandx{\alog}[4][1=\msg,2=q,3=\gname,4=t,usedefault=@]{\big({#1},{#2},{#3},{#4}\big)}

\newcommand{\aQ}{Q}
\newcommandx{\aQzero}[1][1=,usedefault=@]{
  {\ifempty{#1}{q_0}{q_{0#1}}}
}
\newcommand{\badbranches}[1][]{\beta\ifempty{#1}{}{\big({#1}\big)}}
\newcommand{\aTrs}{\tset}
\newcommandx{\guardedaction}[2][1=\al,2=\aguard,usedefault=@]{
  {#1} \ifempty{#2}{}{/} {#2}
}
\newcommandx{\atrM}[4][1=q,2=\al,3={\hat q,\hat \al, \aguard},4=q',usedefault=@]{
  {#1} \xrightarrow[{#3}]{\guardedaction[{#2}][]} {{#4}}
}
\newcommandx{\atrS}[5][
  1={\sysconfig[@][@][\badbranches]},
  2=\al,
  3=\aguard,
  4={\sysconfig[\lstates'][\aconfigfn'][\badbranches]},
  5=\sysctx,usedefault=@
]{
  {#1} \xRightarrow{\qquad} {{#4}}
}
\newcommandx{\arevtrS}[2][
  1={\sysconfig[@][@][\badbranches]},
  2={\sysconfig[\lstates'][\aconfigfn'][\badbranches']},
  usedefault=@
]{
  {#1} \rightsquigarrow {#2}
}

\newcommandx{\enables}[2][1=\aconfigfn,2=\aguard,usedefault=@]{{#1} \vdash {#2}}
\newcommandx{\gprojfn}[5][1=\aG,2=\ptp,3=\cinit,4=\cfinal,5={},usedefault=@]{
  \mathbf{proj}_{#2}({#1},{#3},{#4}\ifempty{#5}{}{,{#5}})
}

\newcommandx{\rbp}[3][1=\aG,2=\aconfigfn,3=\achan,usedefault=@]{\mathtt{RBP}_{{#1},{#2}}\ifempty{#3}{}{\big({#3}\big)}}

\newcommand{\apseudoCFSM}{\mathtt{M}}
\newcommandx{\pseudoseq}[2][1=\apseudoCFSM,2=\apseudoCFSM',usedefault=@]{{#1}  ; {#2}}
\newcommandx{\pseudoCFSM}[4][1=\aQ,2=\aQzero,3=\cfinal,4=\aTrs,usedefault=@]{(#1 \ ; #2 \ ; #3 \ ; #4)}
\newcommandx{\markt}[3][1=\hat{\al},2=\hat{q},3=\aguard,usedefault=@]{\%\big({#1} , {#2}, {#3}\big)}

\newcommandx{\borderfn}[2][1=\aconfig,2=\aloop,usedefault=@]{
  \mathsf{border}_{{#2}}\ifempty{#1}{}{\big({#1}\big)}
}

\newcommandx{\updI}[1]{\mathtt{upd}_\mathtt{I}\ifempty{#1}{}{(#1)}}
\newcommandx{\updO}[1]{\mathtt{upd}_\mathtt{O}\ifempty{#1}{}{(#1)}}

\newcommand{\quo}[1]{\lq\lq {#1}\rq\rq}
\newcommand{\mycomment}[1]{}
\newcommand{\wrt}{\emph{wrt.}\xspace}
\newcommand{\ifempty}[3]{%
  \ifthenelse{\isempty{#1}}{#2}{#3}%
}

\title{Reliability and Fault-Tolerance by Choreographic Design\thanks{Partially supported by EU COST IC1405 (Reversible Computation - Extending Horizons of Computing).}}

\def\titlerunning{Reliability and Fault-Tolerance by Choreographic Design}

\author{
  Ian Cassar\thanks{The research work disclosed in this publication is partially funded by the
  	ENDEAVOUR Scholarships Scheme. ``The scholarship may be part-financed by the
  	European Union --- European Social Fund''}
  \institute{Reykjavik University}\email{ianc@ru.is}
  \and
  Adrian Francalanza
  \institute{University of Malta}\email{adrian.francalanza@um.edu.mt}
  \and  Claudio Antares Mezzina
  \institute{IMT School for Advanced Studies Lucca, Italy}\email{claudio.mezzina@imtlucca.it}
  \and
  Emilio Tuosto
  \institute{University of Leicester, UK}\email{emilio@le.ac.uk}
}
\def\authorrunning{I. Cassar, A. Francalanza, C. A. Mezzina \& E. Tuosto}

\begin{document}

\maketitle              % typeset the title of the contribution

\begin{abstract}
%\mycomment{
  Distributed programs are hard to get right because they are required
  to be open, scalable, long-running, and tolerant to faults.
  In particular, the recent approaches to distributed software based
  on (micro-)services where different services are developed
  independently by disparate teams exacerbate the problem.
  In fact, services are meant to be composed together and run in open
  context where unpredictable behaviours can emerge.
  This makes it necessary to adopt suitable strategies for monitoring
  the execution and incorporate recovery and adaptation mechanisms so
  to make distributed programs more flexible and robust.
  The typical approach that is currently adopted is to embed such
  mechanisms in the program logic, which makes it hard to extract,
  compare and debug.
%}
  %
  We propose an approach that employs formal abstractions for
  specifying failure recovery and adaptation strategies.
  Although implementation agnostic, these abstractions would be
  amenable to algorithmic synthesis of code, monitoring and tests.
  We consider message-passing programs (a la Erlang, Go, or MPI) that
  are gaining momentum both in academia and industry.
  Our research agenda consists of (1) the definition of formal
  behavioural models encompassing failures, (2) the specification of
  the relevant properties of adaptation and recovery strategy, (3) the
  automatic generation of monitoring, recovery, and adaptation logic
  in target languages of interest.
\end{abstract}

% \noindent
% \icnote{Ian, please use this macro for your meta-comments}\\
% \afnote{Adrian, please use this macro for your meta-comments}\\
% \cmnote{Claudio, please use this macro for your meta-comments}

%% Documents sections follow
\section{Introduction}\label{intro:sec}
% !TEX root = main.tex

Distributed applications are notoriously complex and guaranteeing their correctness, robustness, and
resilience is particularly challenging.
These reliability requirements cannot be tackled without considering the problems that
are not generally encountered when developing \emph{non}-distributed
software.
In particular, the execution and behaviour of distributed applications is characterised by a number of factors, a few of which we discuss below:
%
% run in unpredictable execution contexts.
% %
% The factors determining such fragility are many-fold.
%
\begin{itemize}
\item Firstly, communication over networks is subject to failures
  (hardware or software) and to security-related restrictions: nodes may
  crash or undergo management operations, links may fail or be
  temporarily unavailable, access policies may modify the connectivity
  of the system.
\item Secondly, \emph{openness}---a key requirement of distributed
applications---introduces other types of failures.
  A paradigmatic example are (micro-)service architectures where
  distributed components dynamically bind and execute together.
  In this context, failures in the communication infrastructures
  are possibly aggravated by those due to services' unavailability,
  their (behavioural) incompatibility, or to unexpected interactions
  emerging from unforeseen compositions.
\item Also, distributed components may belong to different
  administrative domains; this may introduce unexpected changes to the interaction
  patterns that may not necessarily emerge at design time.
  In addition, unforeseen behaviour may emerge because components may
  evolve independently (e.g., the upgrade of a service may hinder the
  communication with partner services).
\item Another element of concern is that it is hard to determine the
  causes of errors, which in turn complicates efforts to rectify
  and/or mitigate the damage via recovery procedures.
  Since, the boundary of an application are quite \quo{fluid}, it
  becomes infeasible to track and confine errors whenever they emerge.  These
  errors are also hard to reproduce for debugging purposes, and some of them may
  even constitute instances of Heisenbugs \cite{Gray86}.
\end{itemize}

For the above reasons (and others), developers have to harness their
software with mechanisms that ensure (some degree of) dependability.
For instance, the use of monitors capable of detecting failures and
triggering automated countermeasures can avoid catastrophic crashes.
The typical mechanisms used to guarantee fault-tolerance are redundancy
(typically to tackle hardware failures) and exception handling for
software reliability.
It has been observed (see e.g.,~\cite{roo90}) that the use of exception
handling mechanisms naturally leads to defensive approaches in
software development.
For instance, network communications in languages such as Java require
to extensively cast code in try-catch blocks in order to deal with
possible exceptions due to communications.
This muddles the main program logic with auxiliary logic related
to error handling.
Defensive programming, besides being inelegant, is not appealing; in
fact, it requires developers to entangle the application-specific
software with the one related to fault tolerance.

In this position paper, we advocate the use of choreographies to
specify, analyse, and implement reliable strategies for
fault-tolerance and monitoring of distributed message-passing
applications.
We strive towards a setup that teases apart the main program logic
from the coordination of error detection, correction and recovery.
The rest of the paper motivates our approach (\cref{mot:sec}) and
tries to give some hint of the advantages within our choreographic
framework (\cref{cho:sec}).
We draw some conclusions in \cref{conc:sec}.

%%% Local Variables:
%%% mode: latex
%%% TeX-master: "main"
%%% End:

\section{Motivation}\label{mot:sec}
% !TEX root = main.tex

We are interested in \emph{message-passing} frameworks, \emph{i.e.,}
models, systems, and languages where independently executing (distributed) components coordinate
by exchanging messages.
One archetypal model of the message-passing paradigm is the \emph{actor
  model}~\cite{DBLP:books/daglib/0066897} popularised by
industry-strength language implementations such as those found in Akka
(for both Scala and Java)~\cite{Wyatt:2013:AC:2663429},
Elixir~\cite{Thomas:2014:PEF:2723830},
and Erlang~\cite{DBLP:conf/cefp/CesariniT09}.
In particular, one effective approach to fault-tolerance is the model adopted by Erlang.

% To tackle failures,
Rather than trying to achieve absolute error freedom, Erlang's approach
concedes that failures are hard to rule out completely in the setting of
open distributed systems. Accordingly, Erlang-based program development takes into
account the possibility of computation going wrong.  However, instead of
resorting to the usual defensive programming, it adopts
the so-called \quo{let it fail} principle. In place of intertwining
the software realising the application logic with logic for handling errors and
faults, Erlang proposes a supervisory model whereby components
(\emph{i.e.,} actors) are monitored within a
hierarchy of independently-executing \emph{supervisors} (which can monitor for other
supervisors themselves).
When an error occurs within a particular component, it is quarantined by letting that component fail (in isolation); the absence of global shared memory of the actor model facilitates this isolation. Its supervisor is then notified about this failure, creating a traceable event that is useful for debugging. More importantly to our cause, this mechanism also allows the supervisor to take \emph{remedial action} in response to the reported failure.  For instance, the failing component may be restarted by the supervisor. Alternatively, other components that may have been contaminated by the error could also be terminated by the supervisor.  Occasionally supervisors themselves fail in response to a supervised component failing, thus percolating the error to a higher level in the supervision hierarchy.

% according to a recovery policy~\cite{}.

Erlang's model is an instance of a programming paradigm commonly termed as Monitor Oriented Programming (MOP) \cite{meredith-jin-griffith-chen-rosu-2011-jsttt,chen-jin-meredith-rosu-2009-icicis}.  It  neatly separates the application logic from the
recovery policy by encapsulating the logic pertaining to the recovery policy within the supervision structure encasing the application.
Despite this clear advantage, the solution is not without its shortcomings.  For instance, the Erlang supervision mechanisms are still inherently tied to the constructs of the host
language and it is hard to transfer to other technologies. Despite it being localised within supervisor code, manual effort is normally still required to disentangle it from the context where it is defined to be understood in isolation.

We advocate for a recovery mechanism that sits  at a higher level of abstraction than the bare metal of the programming language where it is deployed.
In particular, we envisage the three challenges outlined below:
\begin{enumerate}
\item \label{autom:ch} The explicit identification and design of recovery policies in a technology agnostic manner. This will facilitate the comprehension and understanding of recovery policies and allow for better separation of concerns during program development.
\item \label{monitor:ch} The automated code synthesis from high-level policy descriptions. There only a handful of methods for recovery policy specification and these have limited support for the automatic
  generation of monitors that implement those policies.
  % \afnote{I would say that there is \emph{limited} support.  In some sense,
  % Ian's framework \cite{DBLP:conf/rv/CassarF15,DBLP:conf/ifm/CassarF16} include some of this support already, should one see his
  % scripts as recovery policies.}
\item \label{eval:ch}  The evaluation of recovery policies. We require automated techniques that allow us to ascertain the validity  of recovery policies \emph{w.r.t.} notions of recovery correctness.  We are also  unaware of many frameworks that permit policies to be compared with one another and thus determine whether one recovery policy is better than (or equivalent to) another one.
\end{enumerate}
To the best of our knowledge, there is lack of support to take up the
first challenge.
For instance, Erlang folklore's to recovery policies simply prescribes
the \quo{one-for-one} or the \quo{one-for-all} strategies.
Recently, Neykova and Yoshida have shown how better strategies are
sometimes possible~\cite{ny17}.
We note that the approach followed in~\cite{ny17} is based on simple
yet effective choreographic models.

The second challenge somehow depends on the support one provides
for the design and implementation of recovery strategies.
A basic requirement of (good) abstract software models is that
an artefact (possibly used at different levels of abstractions) has a
clear relationship with the other artefacts that it interacts with, possibly at different levels of abstraction.
This constitutes the essence of model-driven design.
The preservation of these clearly defined interaction points across different abstraction levels is crucial for sound software refinement.  Such a translation from one abstraction level to a more concrete one form the basis for an actual \quo{compilation}
from one model to the other.
%
%  models at an abstraction
% level have clear relations with those at higher or lower
% levels.
%
In cases where such relations have a clear semantics, they can be exploited to
verify properties of the design (and the implementation) as well as to
transform models (semi-)automatically.
In our case, we would expect run-time monitors to be derived from their
abstract models, to ease the development process and allow developer
to focus on the application logic (such as in \cite{DBLP:conf/rv/CassarF15,DBLP:conf/ifm/CassarF16}).

Finally, the right abstraction level should provide the foundations necessary to develop formal techniques to analyse and compare recovery policies as outlined in our third challenge. The right abstraction level would also permit us tractably apply these techniques to specific policy instances; these may either have been developed specifically for the policy formalism considered by the technique or obtained via reverse-engineering methods from a technology-specific application.  Possible examples that may be used as starting points for such an investigation are \cite{DBLP:conf/fossacs/Francalanza16}, where various pre-orders for monitor descriptions are developed, and \cite{DBLP:conf/concur/Francalanza17} where intrinsic monitor correctness criteria such as consistent detections are studied.

% regarding challenge~\ref{eval:ch}, \eMnote{can we say
%   anything here? \cite{ny17} show that this makes sense, but...do we
%   have plans for that?}
%
% We abet the use of choreographies to address the challenges above.
% \eMnote{elaborate on this}

%%% Local Variables:
%%% mode: latex
%%% TeX-master: "main"
%%% End:

\section{Choreographic models and failures }\label{cho:sec}
% !TEX root = main.tex

% We propose a line of research that aims to combine the run-time monitoring and local adaptation of distributed components with the top-down decomposition approach brought about by choreographic development. Our manifesto may thus be distilled as:
% Local Runtime Monitoring + Static Choreography Specifications = Choreographed MOP
% The starting point of our work will rely on two existing bodies of work. Our investigations will, on the one hand, be grounded on the Erlang monitoring framework developed and implemented in [6, 7], (surveyed in Section 3.1) and, on the other hand, be driven by the design for a choreographic model for distributed computation with global views and local projections of [24] (reviewed in Section 3.2).
%

We propose a line of research that aims to combine the run-time
monitoring and local adaptation of distributed components with the top-down decomposition approach brought about by choreographic development.  Our manifesto may thus be distilled as:

\begin{center} \bf
  Local Runtime Adaptation \; + \; Static Choreography Specifications \; = \; Choreographed MOP
\end{center}

% \eMnote{It would be nice to come up with a slogan like monitoring + choreographies = ???
% but what could replace the question marks?
% \cmnote{What about Reliability}
% }
%

The starting point of our work will rely on two existing bodies of work. Our investigations will, on the one hand, be grounded on the Erlang monitoring framework developed and implemented in~\cite{DBLP:conf/rv/CassarF15,DBLP:conf/ifm/CassarF16}, (surveyed in \cref{adpt:sec}) and, on the other hand, be driven by the design of a choreographic model for distributed computation with global views and local projections of~\cite{lty15} (reviewed in \cref{chor:sec}).

% The starting point of our work will be grounded on the monitoring framework for Erlang developed and implemented in~\cite{DBLP:conf/rv/CassarF15,DBLP:conf/ifm/CassarF16},
%  (surveyed in \cref{adpt:sec}) on the one hand, and the choreographic model for distributed computation with global views and local projections of~\cite{lty15}
% (reviewed in \cref{chor:sec}) on the other.

\subsection{A Runtime Adaptation Framework for Erlang Recovery Strategies}
% !TEX root = main.tex

\newcommand{\bnfdef}{\textsf{ ::= }}
\newcommand{\bnfsep}{\; \vert \;}
\newcommand{\bnfsepp}{\!\! \vert \;}

\newcommand{\toolFRM}{\texttt{SPEC}}
\newcommand{\toolFls}{\texttt{flag}}
\newcommand{\toolTru}{\texttt{end}}
\newcommand{\toolNec}[3]{\texttt{[}#1\texttt{]\,rel\,}#2\texttt{.}\,#3}
\newcommand{\toolNecS}[3]{\texttt{*[}#1\texttt{]\,rel\,}#2\texttt{.}\,#3}
\newcommand{\toolAnd}[2]{#1\,\texttt{\&}\,#2}
\newcommand{\toolMax}[2]{\texttt{rec}\,#1\texttt{.}#2\texttt{}}

\newcommand{\toolFall}[2]{\texttt{forall(}#1\texttt{,}#2\texttt{)}}
\newcommand{\toolBool}[2]{\texttt{bool}\,#1\;\texttt{=>}\,#2}
\newcommand{\toolIf}[3]{\texttt{if}\,#1\;\texttt{then}\,#2\,\texttt{else}\,#3}
\newcommand{\toolIfE}[2]{\texttt{if}\,#1\;\texttt{then}\,#2}
\newcommand{\toolAda}[4]{\ensuremath{#1\texttt{(}#2\texttt{)\,rel\,}#3\texttt{.}\,#4}}
\newcommand{\toolAdaG}[3]{{A\texttt{(}#1\texttt{)\,rel\,}#2\texttt{.}\,#3}}
\newcommand{\toolAdaGS}[3]{\ensuremath{S\!\texttt{(}#1\texttt{)\,rel\,}#2\texttt{.}\,#3}}

\newcommand{\toolRestart}[3]{\texttt{restart\!\texttt{(}#1\texttt{)\,rel\,}#2\texttt{.}\,#3}}
\newcommand{\toolFlush}[3]{\texttt{flush\!\texttt{(}#1\texttt{)\,rel\,}#2\texttt{.}\,#3}}

\newcommand{\toolActOutE}[3]{#1\,\texttt{>}\,#2\,\texttt{!}\,#3}
\newcommand{\toolActInE}[3]{#1\,\texttt{<}\,#2\,\texttt{?}\,#3}
\newcommand{\toolActOut}[2]{#1\,\texttt{!}\,#2}
\newcommand{\toolActIn}[2]{#1\,\texttt{?}\,#2}
\newcommand{\toolActCall}[2]{\texttt{call(}#1,#2\texttt{)}}
\newcommand{\toolActRet}[2]{\texttt{ret(}#1,#2\texttt{)}}
\newcommand{\toolTup}[1]{\texttt{\{}#1\texttt{\}}}
\newcommand{\toolList}[1]{\texttt{[}#1\texttt{]}}
\newcommand{\toolN}[1]{\texttt{#1}}

\newcommand{\hVarX}{\texttt{X}}
\newcommand{\hVarY}{\texttt{Y}}
\newcommand{\patt}{\ensuremath{p}}
\newcommand{\pattt}{\ensuremath{q}}
\newcommand{\actC}{\ensuremath{\mu}}
\newcommand{\vLstA}{\ensuremath{\vec{v}}}% [1][]{\ensuremath{r_{#1}}}
\newcommand{\vLstB}{\ensuremath{\vec{u}}}%[1][]{\ensuremath{w_{#1}}}
\newcommand{\vLstI}{\ensuremath{\vec{i}}}
\newcommand{\bV}{\ensuremath{b}}
\newcommand{\bVV}{\ensuremath{b'}}
\newcommand{\adapter}{\textsf{adaptEr}\xspace}
\newcommand{\etal}{\emph{et. al}\xspace}
\newcommand{\eg}{\emph{eg.,}\xspace}
\newcommand{\ie}{\emph{ie.,}\xspace}
\newcommand{\etc}{\emph{etc.}\xspace}

\newcommand{\varN}[1]{\textsl{#1}\xspace}  % variable variable
\newcommand{\xV}{\ensuremath{\varN{x}}\xspace}  % variable variable
\newcommand{\xVV}{\ensuremath{\varN{y}}\xspace}
\newcommand{\xVVV}{\ensuremath{\varN{z}}\xspace}

\newcommand{\vV}{\ensuremath{v}\xspace}  % value variable
\newcommand{\vVV}{\ensuremath{u}\xspace}

\newcommand{\actE}{\ensuremath{\alpha}\xspace}
\newcommand{\actEE}{\ensuremath{\beta}\xspace}

\newcommand{\Pid}{\textsc{Pid}\xspace}
\newcommand{\PAT}{\textsc{Pat}\xspace}
\newcommand{\Vars}{\textsc{Vars}\xspace}
\newcommand{\LVars}{\textsc{Vars}\xspace}
\newcommand{\Bool}{\textsc{Bool}\xspace}
\newcommand{\Val}{\textsc{Val}\xspace}

In \cite{DBLP:conf/rv/CassarF15,DBLP:conf/ifm/CassarF16}, the authors establish local recovery policies via runtime adaptation techniques. 
Runtime adaptation \cite{Kell08survey,Kalareh:phd} is an adaptive monitoring technique prevalent to long-running, highly available software systems, whereby system characteristics  (\eg its structure, locality \etc) are altered \emph{dynamically} in response to runtime events (\eg detected hardware faults or software bugs, changes in system loads), while causing \emph{limited disruption} to the execution of the system.   Numerous examples can be found in service-oriented architectures \cite{Oreizy08SOA,RA-SOA:2008}  (\eg cloud-services, web-services, \etc) for self-configuring, self-optimising and self-healing purposes; the inherent component-based, decoupled organisation of such systems facilitates the implementation of adaptive actions \emph{affecting a subset} of the system while allowing other parts to continue executing normally.

\begin{figure}[h]
	\centering
	\begin{align*}
	c,d \in \toolFRM & \bnfdef \toolFls  && \text{(detect)} && \bnfsepp \toolTru && \text{(terminate)}\\
	& \quad\bnfsepp\;\;\toolAnd{c}{d} && \text{(conjunction)} && \bnfsepp \toolIf{\,\bV\,}{\,c\,}{\,d} && \text{(branch)}\\
	& \quad\bnfsepp\;\;\toolMax{\hVarX}{c} && \text{(recursion)} && \bnfsepp \hVarX && \text{(recursive call)}\\
	& \quad\bnfsepp\;\;\toolNec{\patt}{\vLstA}{c} && \text{(guard)} && \bnfsepp \toolNecS{\patt}{\vLstA}{c} && \text{(blocking guard)} \\
	& \quad\bnfsepp\;\;\toolAdaG{x}{\vLstA}{c} &&\text{(asyn. adaptation)} && \bnfsepp \toolAdaGS{x}{\vLstA}{c} &&\text{(sync. adaptation)}
	\end{align*}
	\caption{Monitor Specification Syntax}
	\label{fig:mon-logic-syntax}
\end{figure}

\paragraph{The Logic} In \cite{DBLP:conf/rv/CassarF15,DBLP:conf/ifm/CassarF16}, Cassar \etal developed the monitoring tool \adapter\footnote{The tool \adapter is open-source and downloadable from \url{https://bitbucket.org/casian/adapter}.} for synthesising adaptation monitors for actor systems developed in Erlang. Specifications in \adapter are defined using the logic in \Cref{fig:mon-logic-syntax}, consisting in a version of Safe Hennessy Milner Logic with recursion (sHML) that is extended with data binding, if statements for inspecting data, adaptations and synchronisation actions.

The logic is defined over streams of visible events, $\actE$,  generated by the monitored system made up of \emph{actors} --- independently-executing processes that are uniquely-identifiable by a process identifier, have their own local memory, and can either spawn other actors or interact with other actors in the system through asynchronous messaging; we use $i,j,h \in \Pid$ to denote the unique identifiers. Monitored events include the sending of messages, $\toolActOutE{i}{j}{v}$, (containing the value $v$ from actor with identifier $i$ to actor $j$), the receipt of messages, $\toolActIn{i}{v}$,  (containing the value $v$ received by actor  $i$), function calls, $\toolActCall{i}{\toolTup{m,f,l}}$, (at actor $i$ for function $f$ in module $m$ with argument list $l$) and function returns, $\toolActRet{i}{\toolTup{m,f,a,v}}$ (at actor $i$ for function $f$ in module $m$ with argument arity $a$ and return value $v$). Event patterns, $\patt,\pattt \in \PAT$ follow a similar structure to that of events, but may contain term variables $\xV,\xVV,\xVVV\in\Vars$  that are bind to concrete values $v,u \in \Val$ (where $\Pid \subseteq \Val$), at runtime through pattern matching (we use $\vec{v}$ to denote lists of values). 

The syntax in \Cref{fig:mon-logic-syntax} includes termination constructs \toolFls\ and \toolTru\ respectively refer to a violation detection and an inconclusive verdict, along with \emph{two} guarding constructs, $\toolNec{\patt}{\vLstI}{c}$ and $\toolNecS{\patt}{\vLstI}{c}$, instructing the respective monitor to observe system events that match event pattern \patt, and progressing as $c$ if the match is successful. These constructs encompass directives for \emph{blocking} and \emph{releasing} actor executions, depending on the events observed. The guarding construct $\toolNecS{\patt}{\vLstI}{c}$ is \emph{blocking}, meaning that it \emph{suspends} the execution the actor whose identifier is the \emph{subject} of the event matched by the pattern (\eg actor $i$ is the subject in the events $\toolActOutE{i}{j}{v}$,  $\toolActInE{i}{j}{v}$, $\toolActCall{i}{\toolTup{m,f,l}}$ and $\toolActRet{i}{\toolTup{m,f,a,r}}$). By contrast, the  guarding construct $\toolNec{\patt}{\vLstI}{c}$ does not block any actor when its pattern is matched.  However, for both  constructs $\toolNec{\patt}{\vLstI}{c}$ and $\toolNecS{\patt}{\vLstI}{c}$, pattern \emph{mismatch} terminates monitoring, and \emph{also} releases all the blocked actors in the list of identifiers \vLstI. 
In addition to term variables, the abstract syntax in \Cref{fig:mon-logic-syntax} also assumes a distinct denumerable set of \emph{formula variables} $\hVarX,\hVarY,\ldots\in\LVars$, used to define recursive specifications.     It is also parametrised by a set of \emph{decidable} boolean expressions, $\bV,\bVV\in\Bool$, and the aforementioned set of event patterns. Monitor specifications include commands for flagging violations, $\toolFls$, and terminating (silently), $\toolTru$, conjunctions, $\toolAnd{c_1}{c_2}$, recursion, $\toolMax{\hVarX}{c}$, and conditionals to reason about data, $\toolIf{\bV}{c_1}{c_2}$ --- we encode $\toolIfE{\bV}{c_1}$ as $\toolIf{\bV}{c_1}{\toolTru}$.

The syntax in \Cref{fig:mon-logic-syntax} also specifies two adaptation constructs, $\toolAdaG{j}{\vLstI}{c}$ and $\toolAdaGS{j}{\vLstI}{c}$. Both constructs instruct the monitor to administer an adaptation action ($A$ and $S$) on actor $j$, releasing the (blocked) actors in the list \vLstI\ afterwards,  then progressing as $c$.  The only difference between these two constructs is that the adaptation in $\toolAdaGS{j}{\vLstI}{c}$, namely $S$, expects the target actor $j$ to be \emph{blocked} (\ie synchronised with the monitor) when the adaptation is administered, and must therefore be blocked by some preceding guarding construct. Synchronous adaptations include, amongst others (see \cite{DBLP:conf/ifm/CassarF16,DBLP:conf/rv/CassarF15}), $\toolRestart{x}{\vLstA}{c}$ and $\toolFlush{x}{\vLstA}{c}$ for restarting and emptying the mailboxes of misbehaving actors.

\begin{figure}[t]% [htbp]
	\centering
	\begin{tikzpicture}[>=latex,auto,thick]
	\begin{scope}[draw=blue!50,fill=blue!20,minimum size=0.60cm, align=left]
	\node (Incrementor) at (-2,0) [shape=rectangle,draw,fill]  {\;\,\quad\textbf{Incrementor}\quad};
	\node (Decrementor) at (2,0) [shape=rectangle,draw,fill]  {\;\quad\textbf{Decrementor}\quad};
	\node (Interface) at (0,-1.35) [shape=rectangle,draw,fill]  {\;\quad\textbf{Common Interface}\quad};
	\end{scope}

	\begin{scope}[draw=black,fill=black, align=right, font=\small]
	\node at ([xshift=0.23cm,yshift=-0.25cm]Incrementor.north west) [shape=rectangle,draw,fill,font=\small]  {${\color{white}j}$ };
	\node at ([xshift=0.20cm,yshift=-0.23cm]Decrementor.north west) [shape=rectangle,draw,fill,font=\small]  {${\color{white}h}$\! };
	\node  at ([xshift=0.21cm,yshift=-0.22cm]Interface.north west) [shape=rectangle,draw,fill,font=\small]  {${\color{white}i}$ };
	\end{scope}

	\begin{scope}[align=center]
	\node (in) at (0,-2.4)  {(1) $\toolTup{\toolN{inc},3,cli}$};
	\node (outGood) at (-3.5,-2.4)  {$(3)\;\toolTup{\toolN{res},4}$};
	\node (outErr) at (3.5,-2.4)  {$\color{red}(3)\;\toolN{err}$};
	\end{scope}
	\begin{scope}[draw=black]
	\draw[% dashed,
	->] (in) [align=center] to node [left]{} (Interface);
	\draw[->] (Interface) [align=center,bend right=25] to node [left]{\\[1mm] $(2)\;\toolTup{\toolN{inc},3,cli}$} (Incrementor);
	\draw[->] (Incrementor) [align=center,bend right=25] to node [left]{} (outGood);
	% \draw[dashed,-] (4,-2.) to (-4,-2.);
	\end{scope}
	\begin{scope}[draw=blue]
	\draw[dashed,-] (4,-1.9) to (-4,-1.9);
	\node (ext)  at (-5,-1.9) {\textcolor{blue}{External View}};
	\end{scope}
	\begin{scope}[draw=red, fill=red]
	\draw[->] (Interface) [align=center,bend left=25] to node [right]{\\[1mm] $\color{red}(2)\;\toolTup{\toolN{inc},3,cli}$} (Decrementor);
	\draw[->] (Decrementor) [align=center,bend left=25] to node [right]{} (outErr);
	\end{scope}
	\end{tikzpicture}
	\caption{A server actor implementation offering integer increment and decrement services}
	\label{fig:sys}
	\vspace{-2mm}
\end{figure}
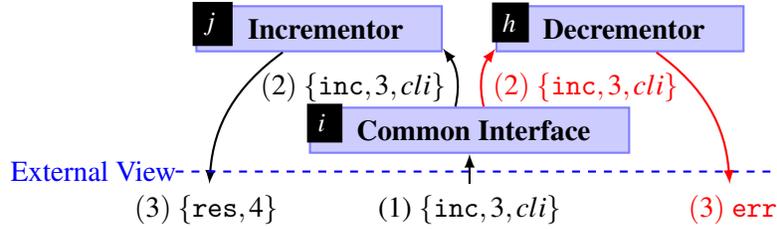

\begin{example} \normalfont
	\label{ex:intro}
	\Cref{fig:sys} depicts a server  consisting of % $(i)$
	a front-end \emph{Common Interface} actor with identifier $i$ receiving client requests, % $(ii)$
	a back-end \emph{Incrementor} actor with identifier $j$, handling integer increment requests, and % $(iii)$
	a back-end \emph{Decrementor} actor  $h$, handing decrement requests.  A client sends service requests to actor $i$ of the form $\toolTup{\textsl{tag},\textsl{arg},\textsl{ret}}$ where \textsl{tag} selects the type of service, \textsl{arg} carries the service arguments and \textsl{ret} specifies the return address for the result % is to be sent
	(typically the client actor ID). The interface actor forwards the request to one of its back-end servers (depending on the \textsl{tag}) whereas the back-end servers process the requests, sending results (or error messages) to \textsl{ret}. The tool \adapter allows us to specify safety properties such as \eqref{eq:1}, explained below:
	\begin{align}
	\!\!\!\label{eq:1}
	& \toolMax{\,\hVarY}{\toolNecS{\toolActOutE{i}{\_}{\toolTup{\toolN{inc},x,y}}}{\epsilon}{}\left(\!\begin{array}{lll}
	 (\,\toolNec{\toolActOutE{j}{y}{\toolTup{\toolN{res},\smash{x+1}}}}{\epsilon}{\hVarY}) \;\toolAnd \;  \\[0mm]
		 (\,\toolNecS{\toolActOutE{z\,}{y}{\toolN{err}}}{\epsilon}{\; \toolIfE{z\in\{j,h\}}{\\\qquad\toolRestart{i}{$\epsilon$}{}\toolFlush{z}{\toolList{i,z}}{\hVarY}}})
			\end{array}\right)
	}
	\end{align}
	It is a (recursive, \toolMax{\,\hVarY}{\ldots} )  property requiring that, from an \emph{external} viewpoint,  \emph{every} increment request sent by $i$ to either $j$ or $h$, action  $\toolActOutE{i}{\_}{\toolTup{\toolN{inc},x,y}}$,  is followed by an answer from  $j$ to the  address $y$ carrying $x+1$,  action $\toolActOutE{j\,}{\,y}{\toolTup{\toolN{res},\smash{x+1}}}$ (recursing through variable \hVarY). However, increment requests followed by an error message sent from \emph{any} actor back to $y$, action $\toolActInE{\_\,}{y}{\toolN{err}}$,  represent a violation which requires mitigation. To enable mitigation through \emph{synchronous adaptations} it is crucial to block the respective actors beforehand, in case they contribute to a property violation. We therefore specify blocking guard $\toolNecS{\toolActOutE{i}{\_}{\toolTup{\toolN{inc},x,y}}}{\epsilon}{}$ to block actor $i$ upon forwarding the event to any one of the backend actors (\ie either $j$ or $h$), along with $\toolNecS{\toolActOutE{z\,}{y}{\toolN{err}}}{\epsilon}$ to block any $z$ actor that generates an error. Before mitigating this error, the monitor inspects the value bound to data variable $z$ by using an if-statement; this ensures that the adaptations are being performed on the back-end actors.

	To mitigate the violation we then specify that the monitor should restart actor $i$ with synchronous adaptation \toolRestart{i}{$\epsilon$}{}, and empty the mailbox of the back-end server---which may contain more erroneously forwarded messages---through adaptation \toolFlush{z}{\toolList{i,z}}{} (the actor to be purged is determined at runtime, where $z$ is bound to identifier $h$ from the previous action $\toolNec{\toolActInE{z\,}{y}{\toolN{err}}}{\epsilon}$).  Importantly, note that in the above execution (where $h$ is the actor sending the error message), actor $j$ is \emph{not affected} by any adaptation action taken. The last adaptation also contains pids $i$ and $z$ in its release list; this ensures that whenever the received input number $x$ is correctly incremented, these blocked actors are released before the monitor recurs

\end{example}

%\paragraph{The Implementation}

%%% Local Variables:
%%% mode: latex
%%% TeX-master: "main"
%%% End:
\label{adpt:sec}

\subsection{Global and Local Specifications}
% !TEX root = main.tex

A key reason that makes choreographies appealing for the modelling,
design, and analysis of distributed applications is that they do not
envisage centralisation points.
Roughly, in a choreographic model one describes how a few distributed
components interact in order to coordinate with each other.
There is a range of possible interpretations of
choreographies \cite{bdft16}; a widely accepted informal description
is the one suggested by W3C's~\cite{w3c:cho}:
\begin{quote}\footnotesize
  [...] %Using the Web Services Choreography specification,
  a contract containing a global definition of the common ordering
  conditions and constraints under which messages are exchanged, is
  produced that describes, from a \textbf{global viewpoint} [...]
  observable behaviour [...]. Each party can then use the
  \textbf{global definition} to build and test solutions that conform
  to it. The global specification is in turn realised by combination
  of the resulting \textbf{local systems} [...]
\end{quote}
According to this description, a \textbf{global} and a \textbf{local}
views are related as in the left-most diagram in \cref{fig:global-local-adapt-straregies}
which evokes the following software engineering approach.
First, an architect designs the global specification; then the
architect uses the global specification to derive, via a \lq
projection\rq\ operation, a local specification for the distributed
components the local ones; finally, programmers can use the local
specifications to check that the implementation of their components
are compliant with the local specification.
The keystones of this process are ($i$) that the global specification
can be used to guarantee good behaviour of the system abstracting away
from low level details (typically assuming synchronous
communications), ($ii$) that projection operation can usually be
automatised so to ($iii$) produce local specifications at a lower
level of abstraction (where communication are asynchronous) while
preserving the behaviour of the global specification.

We remark that the relations among views and systems of choreographies
are richer than those discussed here.
For instance, local views can also be compiled into template code of
components and the projection operation may have an \quo{inverse}
(cf.~\cite{lty15}).
Those aspects are not in our scope here.

We choose two specific formalisms for global and local specifications.
More precisely, we adapt to our needs the \emph{global graphs}
of~\cite{lty15} for global specifications and \emph{communicating
  finite-state machines} (CFSMs)~\cite{bz83} to formalise local views
of choreographies.

\begin{figure}[ht]
  \centering
  $\begin{array}{c@{\qquad}c@{\qquad}c@{\qquad}c@{\qquad}c@{\qquad}c}
     % gint
     \begin{tikzpicture}[node distance=0.4cm and 0.4cm, every node/.style={scale=.5,transform shape}]
       \node[src] at (0,2) (src) {};
       \node[block] at (0,1) (int) {$\gint[{}]$};
       \node[sink] at (0,0) (sink) {};
       \path[line] (src) -- (int);
       \path[line] (int) -- (sink);
     \end{tikzpicture}
     &
     % gseq
     \begin{tikzpicture}[node distance=0.7cm and 0.4cm, every node/.style={scale=.5,transform shape}]
       \node[bblock] at (0,0) (g)
       {$\aG$}; \node[node, below=of
       g] (s1) {}; \node[bblock, below=of s1]
       (gp)
       {$\aG'$}; \path[line,dotted] (g) -- (s1); \path[line,dotted]
       (s1) -- (gp);
     \end{tikzpicture}
     &
     % gpar
     \begin{tikzpicture}[node distance=0.4cm and 0.4cm, every node/.style={scale=.5,transform shape}]
       \node[src] at (0,0) (src) {};
       \node[agate,below=of src] (par) {};
       \node[node, left=of par] (s1) {};
       \node[node, right=of par] (s2) {};
       \node[bblock, below=of s1] (g) {$\aG_1$};
       \node[bblock, below=of s2] (gp) {$\aG_h$};
       \node[node, below=of g] (sg) {};
       \node[node, below=of gp] (sgp) {};
       \node[agate, right=of sg,label=above:{}] (join) {};
       \node[sink, below=of join] (sink) {};
       \path[line] (src) -- (par);
       \path[line] (par) -- (s1);
       \path[line] (par) -- (s2);
       \path[line,dotted] (s1) -- (g); \path[line,dotted] (s2) -- (gp);
       \path[line,dotted] (g) -- (sg); \path[line,dotted] (gp) -- (sgp);
       \path[line] (sg) -- (join); \path[line] (sgp) -- (join);
       \path[line] (join) -- (sink);
     \end{tikzpicture}
     &
     % grec
     \begin{tikzpicture}[node distance=0.4cm and 0.4cm, every node/.style={scale=.5,transform shape},auto]
       \node[src] at (0,2) (src) {};
       \node[ogate,below=of src] (loopentry) {}; 
       \node[bblock,below=of loopentry] (body) {$\aG$};
       \node[ogate,below=of body] (loop) {}; 
       \node[sink,below=of loop] (sink) {};
       \path[line,dotted] (src) -- (loopentry);
       \path[line] (loopentry) -- (body);
       \path[line] (loop) -- (sink);
       \path[line,dotted] (body) -- (loop);
       \path[line] (loop) edge [bend right=45] node {} (loopentry);
     \end{tikzpicture}
     &
     % gcho
     \begin{tikzpicture}[auto,node distance=0.4cm and 0.4cm, every node/.style={scale=.5,transform shape}]
       \node[src] at (0,0) (src) {};
       \node[ogate,below=of src] (par) {};
       \node[node, left=of par] (s1) {};
       \node[node, right=of par] (s2) {};
       \node[bblock, below=of s1] (g) {$\aG_1$};
       \node[bblock, below=of s2] (gp) {$\aG_h$};
       \node[node, below=of g] (sg) {};
       \node[node, below=of gp] (sgp) {};
       \node[ogate, right=of sg] (join) {};
       \node[sink, below=of join] (sink) {};
       \path[line] (src) -- (par);
       \draw[line] (par) to node[swap] {} (s1);
       \draw[line] (par) to node {} (s2);
       \path[line,dotted] (s1) -- (g);
       \path[line,dotted] (s2) -- (gp);
       \path[line,dotted] (g) -- (sg);
       \path[line,dotted] (gp) -- (sgp);
       \path[line,dotted] (sg) -- (join);
       \path[line,dotted] (sgp) -- (join);
       \path[line] (join) -- (sink);
     \end{tikzpicture}
     &
       \begin{tikzpicture}
         \node[draw] (text) {
           \begin{minipage}[c]{.3\linewidth} \tiny
             A global graph $\aG$ is a rooted graph with a single
             \quo{enter} and a single \quo{exit} nodes, called
             \emph{source} and \emph{sink} respectively and depicted
             as $\gsource$ and $\gsink$.
             Other nodes are drawn as $\gvertex$ while dotted edges
             have a \quo{meta-meaning} to single out source and sink
             nodes of sub-graphs.
             More precisely, a dotted edge from a node $\gvertex$ to a
             boxed $\aG$ means that $\gvertex$ is the source of $\aG$;
             similarly, a dotted edge from a boxed $\aG$ to $\gvertex$
             means that $\gvertex$ is the sink of $\aG$.
             For instance, the top-most dotted edge for the sequential
             composition identifies the sink node of $\aG$ while the
             other dotted edge identifies the source node of $\aG'$;
             in other words, the sequential composition of $\aG$ and
             $\aG'$ is obtained by coalescing the sink of $\aG$ with
             the source of $\aG'$.
           \end{minipage}
};
       \end{tikzpicture}
     \\
     \text{\small interaction}
     &
     \text{\small sequential}
     &
     \text{\small parallel}
     &
     \text{\small iteration}
     &
     \text{\small branching}
   \end{array}$
   \caption{A graphical notation for choreographies}
   \label{fig:graphs}
\end{figure}
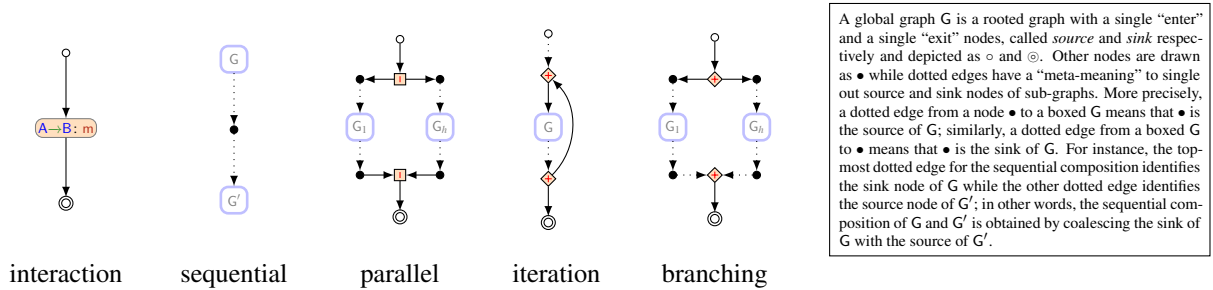

\paragraph{Global Specifications}
Global graphs can be seen as a graphical model of distributed
work-flows and communicating machines (a well established model for
communication protocol design) fix the distribution and communication
model.
We describe global graphs using their graphical notation reported in
\cref{fig:graphs} featuring simple \textit{interactions} (e.g.,
party $\p$ sending a message $\msg$ to party $\q$),
\textit{sequential} or \textit{parallel} compositions of global
graphs, \textit{iteration} of global graphs, or \textit{choice}
between alternative global graphs.
%
%We now explain the semantics of global graphs.

\mycomment{
A global graph $\aG$ is a rooted graph with a single \quo{enter} and a
single \quo{exit} control points called \emph{source} and \emph{sink}
respectively.
There, nodes $\gsource$ and $\gsink$ respectively denote the source
node the sink node; other nodes are drawn as $\gvertex$ while dotted
edges have a \quo{meta-meaning} to single out source and sink nodes.
More precisely, a dotted edge from a node $\gvertex$ to a boxed $\aG$
means that $\gvertex$ is the source of $\aG$; similarly, a dotted edge
from a boxed $\aG$ to $\gvertex$ means that $\gvertex$ is the sink of
$\aG$.
For instance, the top-most dotted edge for the sequential composition
in \cref{fig:graphs} identifies the sink node of $\aG$ while the other
dotted edge identifies the source node of $\aG'$; in other words, the
sequential composition of $\aG$ and $\aG'$ is obtained by coalescing
the sink of $\aG$ with the source of $\aG'$.

\newcommand{\traveller}{\p[A]}
\newcommand{\broker}{\q}
\newcommand{\db}{\p[d]}
\newcommand{\exbound}{1}
\newcommand{\exseqAB}[4]{\gseq[][{\gint[{#1}][\traveller][{\msg[{#2}]}][\broker]}][{\gint[{#3}][\broker][{\msg[{#4}]}][\traveller]}]}

%\begin{wrapfigure}{r}{0.5\textwidth}
 %    \includegraphics[width=0.48\textwidth]{gull}
%  \end{center}
%  \caption{A gull}
%\end{wrapfigure}
  \begin{tikzpicture}[node distance=0.6cm and 0.4cm, every
    node/.style={scale=.45,transform shape}]
    \node[src] at (0,2) (src) {};
    \node[ogate,below= of src,label=left:{$\gname[1]$}] (loopentry) {}; 
    \node[ogate,below= of loopentry,label=below:{$\gname[2]$},label={[xshift=-2cm]
},label={[xshift=2cm]}] (branchstart) {};
    \node[agate, below left= of branchstart,label=below:{$\gname[3]$},xshift=-1.5cm] (fork) {};
    \node[block, below left = of fork,label=left:{$\gname[4]$} ] (flight) {$\gint[{}][\traveller][{\msg[restaurant]}][\broker]$};
    \node[block, below  = of flight,label=left:{$\gname[5]$} ] (flightprice) {$\gint[{}][\broker][{\msg[restPrice]}][\traveller]$};
    \node[block, below right=of fork,label=left:{$\gname[6]$} ] (car) {$\gint[{}][\traveller][{\msg[movie]}][\broker]$};
    \node[block, below = of car,label=left:{$\gname[7]$} ] (carprice) {$\gint[{}][\broker][{\msg[moviePrice]}][\traveller]$};
    \node[block, right = of car,label=right:{$\gname[8]$},xshift=.5cm] (dest) {$\gint[{}][\traveller][{\msg[date]}][\broker]$};
    \node[block, below = of dest, label=right:{$\gname[9]$},xshift=.1cm] (destprice) {$\gint[{}][\broker][{\msg[fullPrice]}][\traveller]$};
    \node[agate, below left = of carprice, label=above:{$\nmerge{\gname[3]}$}] (join) {};
    \node[ogate, below right = of join, label=above:{$\nmerge{\gname[2]}$}] (branchstop) {};
    \node[block, below = of branchstop,label=right:{$\gname[10]$}] (upd) {$\gint[{}][{\traveller}][{\msg[upd]}][\db]$};
    \node[ogate, below = of upd, label=left:{$\nmerge{\gname[1]}$}] (loopexit) {};
    \node[sink, below = of loopexit] (sink) {};
    \node[right = of loopexit,xshift=5cm] (dummy) {};
    \path[line] (src) -- (loopentry);
    \path[line] (loopentry) -- (branchstart);
    \path[line] (flight) -- (flightprice);
    \path[line] (car) -- (carprice);
    \path[line] (dest) -- (destprice);
    \path[line] (branchstart) -| (fork);
    \path[line] (branchstart) -| (dest);
    \path[line] (fork) -| (flight);
    \path[line] (fork) -| (car);
    \path[line] (flightprice) |- (join);
    \path[line] (carprice) |- (join);
    \path[line] (join) |- (branchstop);
    \path[line] (destprice) |- (branchstop);
    \path[line] (branchstop) -- (upd);
    \path[line] (upd) -- (loopexit);
    \path[line] (loopexit.east) -- (dummy) |- (loopentry.east);
    \path[line] (loopexit) -- (sink);
  \end{tikzpicture}
%\end{wrapfigure}
}
The semantics of a global graph $\aG$ is given in terms of
partial-orders on the communication \emph{events} of $\aG$.
For space limitations, we can only give an intuitive idea of such
semantics (see~\cite{gt16} for details).
The semantics of an interaction is straightforward: the output event
of the sender must precede the input event of the receiver.
Likewise, the semantics of parallel composition is obvious: the events
of a thread $\aG_i$ do not have any order relation with those of a
thread $\aG_j$ with $i \neq j$.
The other cases are more delicate.
For iteration, we take the unfolding of the loop and propagate the
order accordingly (this is similar to what done with processes of
Petri nets).
The semantics of $\gseq[{}]$ is defined provided that (i) the
semantics of $\aG$ and $\aG'$ are defined, (ii) each input event in
$\aG'$ follows each output event in $\aG$ in the order built by taking
the reflexo-transitive closure of union of semantics of $\aG$ and of
$\aG'$ after adding the dependencies between the events in $\aG$ and
those in $\aG'$ with the same subject.
The semantics of a choice is defined provided that the semantics of
each branch $\aG_i$ is defined and that the choice is
\emph{well-branched}, namely that there is only one party deciding
which branch to take and that all the other parties involved in the
choice can discriminate using their input events.

\paragraph{Local Specifications}
We adopt systems of CFSMs~\cite{bz83} as our 
model of local specifications.
A CFSM is a finite-state automaton where transitions represent input
or output events from/to other machines.
Each machine in the system corresponds to an actor which can send or
receive messages to/from other machines.
Communications take place on unbound FIFO buffers: for each pair of
machines, say $\p$ and $\q$, there is a buffer from $\p$ to $\q$ and
one from $\q$ to $\p$.
Basically, when a machine $\p$ is in a state $q$ with a transition to
a state $q'$ whose label is an output of message $\msg$ to $\q$, then
$\msg$ is put in the buffer from $\p$ to $\q$ and $\p$ moves to state
$q'$.
Similarly, when $\q$ is in a state $q$ with a transition to a state
$q'$ whose label is an input of $\msg$ from $\q$ and the $\msg$ is on
the top of the buffer from $\p$ to $\q$ then $\q$ pops $\msg$ from the
buffer and moves to state $q'$.

Noteworthy, the model of CFSMs is very close to the actor model and
CFSMs can be projected from global graphs automatically.
Moreover, when the global graph, say $\aG$, is \emph{well-formed} then
the behaviour of the projected machines faithfully refines the
semantics of $\aG$~\cite{gt16}.

%%% Local Variables:
%%% mode: latex
%%% TeX-master: "main"
%%% End:
\label{chor:sec}

%\eMnote{RA or monitoring? Make this consistent}
%\eMnote{Can we remove \cref{fig:mon-logic-syntax} and explain the syntax through the example?}

\section{The Proposed Approach}\label{agenda:sec}
% !TEX root = main.tex

\begin{figure}[h]
	\centering
% \begin{center}
\includegraphics[scale=.35]{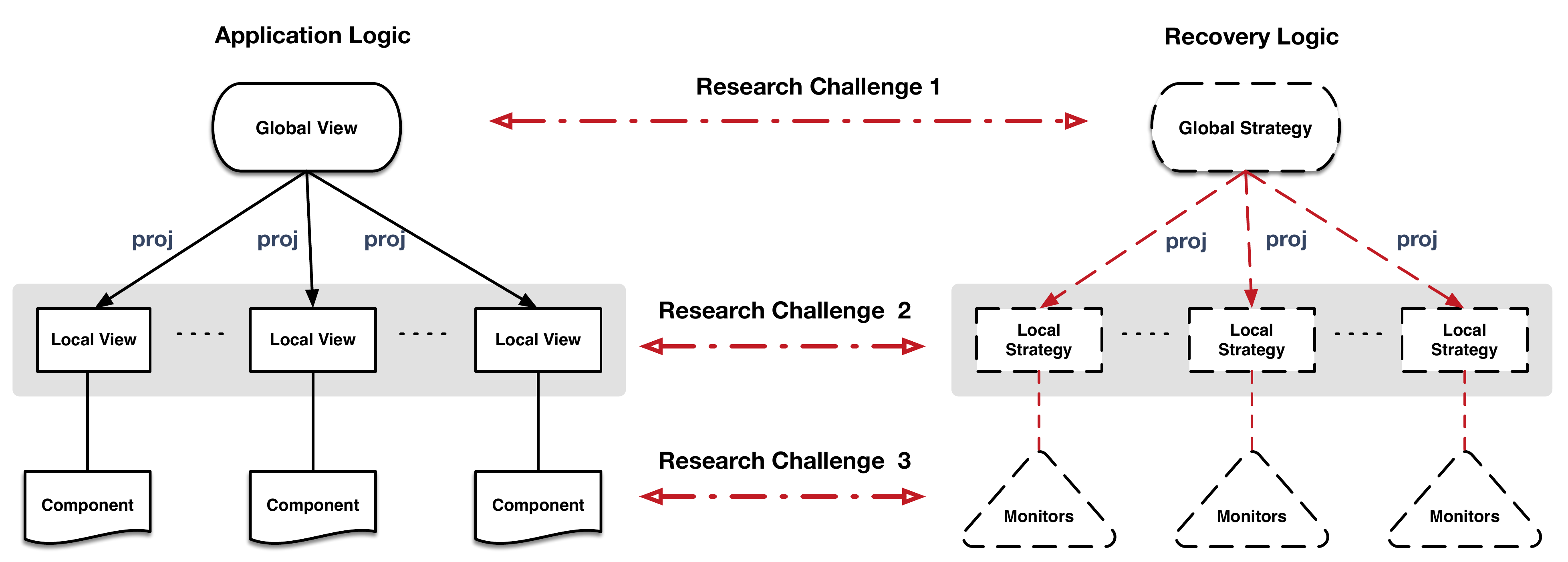}
% \end{center}
\caption{A Global-Local approach to Adaptation Strategies}
\label{fig:global-local-adapt-straregies}
\end{figure}

We advocate that the development of recovery logic is
\textit{orthogonal} to the application logic, and
this separation of concerns could induce separate development
efforts which are, to a certain degree, independent from one another.
% in parallel.
%
Similar to the case for the application logic, we envisage global and local points of
view for the recovery logic whereby the latter  is attained by projecting
the global strategy.
Our approach is schematically described in Figure~\ref{fig:global-local-adapt-straregies}.
The left-most part of the diagram illustrates the top-down approach of
choreographies of the application logic described in \cref{chor:sec}.
We propose to develop a similar approach for the recovery logic as
depicted in the right-most part of
\cref{fig:global-local-adapt-straregies}, where the triangular shape
for monitors evokes that monitors are possibly arranged in a complex
structure (as e.g., the \emph{hierarchy} of Erlang supervisors).
In fact, we envisage that a local strategy could correspond to a
subsystem of monitors as in the case of \cite{DBLP:conf/ifm/CassarF16,DBLP:conf/rv/AttardF16} (unlike the choreographies for the application
logic, where each local view typically yields one component).

\newcommand{\toolRoll}[3]{\texttt{roll\!\texttt{(}#1\texttt{)\,rel\,}#2\texttt{.}\,#3}}

\paragraph*{Models to express Global and Local Strategies.}
Choreographic models have to be equipped with features allowing us to
design and analyse the recovery logic of systems.
This requires, on the one hand, the identification of suitable
linguistic mechanisms for expressing global/local strategies and, on
the other hand, to define principle of monitors programming by looking
at state-of-the-art techniques.
For example, the (global) recovery logic could allow us to specify
\emph{recovery} points where parties can roll-back if some kind of
error is met or \emph{compensations} to activate when anomalous
configurations are reached.

A challenge here is the definition of projection operations that
enable featuring recovery mechanisms.
A first step in this direction is a recent proposal of Mezzina and
Tuosto~\cite{corr/MezzinaT17} who extend the global graphs reviewed in
\cref{chor:sec} with \emph{reversibility guards} to recover the system
when it reaches undesired configurations.
A promising research direction in this respect is to extend the language
of reversibility guards with the patterns featured by \adapter\ and
then define projection operations to automatically obtain \adapter\ monitors.

\newcommand{\enc}[1]{\textsf{enc}(#1)}
For example, consider the following scenario in which a party $\ptp[A]$ is required to plan a trip where he may choose to take a flight or to travel via train. Depending on this choice his behaviour differs given that if he books a flight, he must also book a taxi for the date, while if he takes the train he can manage to schedule a meeting for the very same day since the station is close-by. Moreover, if he chooses to take the flight but is then unable to book the flight, he must then revert back to the train option. Following the approach of~\cite{corr/MezzinaT17} this can be expressed by the following (global) choreography:  

%$\gint[{}][@][m_1][{\ptp[C]}]$
$$\gcho[{}][{\agG[{\gseq[{}][{\gint[{}][@][FlightAndTaxi]}][\aG_1]}][\aguard_1]}][{\agG[{\gseq[{}][{\gint[{}][@][Train][{\ptp[C]}]}][\aG_2]}][\colorGuard{f\!\! f}]}]$$
where $\aG_1$ and $\aG_2$ describe the global behaviour for the system, followed by the choice guard $\aguard_1$ which expresses the fact that the taxi is unavailable; $\colorGuard{f\!\!f}$ represents the falsehood condition. Starting from this global choreography, the appropriate local strategies must be derived accordingly. For instance, one way to go about this is by creating an encoding function $\enc{\aguard_1}$ that allows for redefining choice guards (such as $\aguard_1$) in terms of \adapter's logic. In this way the designated monitor can then verify the entire execution branch \wrt the encoded version of $\aG$, \ie $\enc{\aguard_1}$.Alternatively, we can also extend \adapter's logic to allow for defining choreography guards as patterns within \adapter's guards in the lines of $\toolNecS{\aguard}{\vLstA}{c}$.

Furthermore, support for additional adaptation constructs such as $\toolRoll{i}{\vLstA}{}$ must be added in order to augment reversibility features within \adapter. Such an adaptation should permit the monitor to reverse specific parts of the computation of an offending actor $i$ along with the computation of other actors which had interacted with $i$. For instance, a local strategy for $\ptp[A]$, while executing the left branch of the choice, can be defined in terms of \adapter's logic as follows:
\begin{align*}
& \toolMax{\,\hVarY}{\toolNecS{\textsf{act}(\ptp[A])}{\epsilon}{}\left(\!\begin{array}{lll}
	(\,\enc{\neg \aguard_1}\toolRoll{\ptp[A]}{\ptp[A]}{\hVarY}) \\ \toolAnd 
	\,(\enc{\aguard_1}{\hVarY}) 
	\end{array}\right)
}
\end{align*}
The above script states that every time $\ptp[A]$ performs some action $\textsf{act}(A)$ (regardless of the type of action), this is blocked such that whenever the encoded version of $\aguard_1$ (\ie $\enc{\aguard_1}$) evaluates to false, a reversing adaptation is applied to $\ptp[A]$ via $\toolRoll{A}{A}{}$ which then releases $\ptp[A]$; otherwise, the computation of $\ptp[A]$ is left intact since the monitor releases $\ptp[A]$ immediately. 
Naturally the reversing adaptation has to revert all the actions of $\ptp[A]$ and to notify the parties involved in these actions. 

\paragraph*{Properties of Recovery Logic.}
\mycomment{ Standard properties of inter the application model of
  global/local views enforces are \emph{deadlock freedom} and absence
  of orphan messages.
  On the other side, we}
We have to understand general properties of interest of recovery
as well as specific ones.
One general property could be the fact that the strategy guides the
application toward a \textit{safe} state when errors occur.
For example, the recovery strategy could guarantee \emph{causal
  consistency}, namely that a safe state is one that the execution
reached previously.
Recovery strategies may be subject to resource requirements that need to be
taken into consideration and/or adhered to.  One such example would be the
% Other examples of properties could be:
minimisation of the number of components that have to be re-started
when a recovery procedure is administered, whereby the restarted
components are causally related to the error detected.
% the
% minimisation of computational steps that component have to undertake
% with a recovery strategy, that is instead of restarting from scratch a
% component could just revert some of its computation; the minimisation
% of coordination messages among monitors.
%
The work discussed in Section~\ref{adpt:sec} provides another example of resource
requirements for recovery strategies: in an asynchronous monitoring setting,
component synchronisations are considered to be expensive operations and,
as a result, the monitors are expected to use the least number of
component synchronisations for the adaptation actions to be
administered correctly.

Also, as typical for choreographies, we should unveil the conditions
under which a recovery strategy is realisable in a distributed settings.
In other words, not all globally-specified recovery policies are necessarily
implementable in a choreographed distributed setting; we therefore seek to
establish \textit{well-formedness}  criteria that allow us to determine when a
global recovery policy can be projected (and thus implemented) in a
decentralised setup.

\paragraph*{Compliance.} In the case of recovery strategies, it is unclear
when monitors are deemed to be compliant with their local strategy.
A central aspect that we need to tackle is that of understanding what
it actually means for monitors and local strategy to be compliant, and
subsequently to give a suitable compliance definition that captures this
understanding.
One possible approach to address this problem is to emulate and extend what was done for
the application logic where  several notions of behavioural compliance
have been studies (\eg \cite{DBLP:journals/toplas/CastagnaGP09,DBLP:journals/corr/BernardiH15}).
%
% However, we do not expect these theories to carry over in a straightforward fashion because of a number of  the fact that monitors may
% be organised in complex structures (e.g., hierarchies of monitors).
%

Another potential avenue worth considering is the work on monitorability \cite{FraAI17}
that relates the behaviour of the monitor to that specified by the correctness property of interest;
the work in \cite{FraSey14} investigates these issues for a target actor calculus that is deeply inspired by the Erlang model.
In such cases we would need to extend the concept of monitorability to adaptability and enforceability
with respect to the local strategy derived from the global specification.
Once we identify and formalise our notions of compliance, we then need to study their decidability
properties, and investigate approaches to check compliance such as
type-checking or behavioural equivalence checking (\emph{e.g.,} via testing preorders or
bisimulations).

\paragraph*{Seamless Integration.}
A key driving principle of our proposed programme is that the recovery logic
has to be orthogonal to the application logic.
This separation of concerns would allow the traditional designers to focus on the
application logic and just declare the error conditions to be managed
by the recovery logic.
The dedicated designers of the recovery logic would then use those error conditions
and the structure of the choreography of the application logic to
specify a recovery strategy.
Finally, the application and recovery logic will have to be integrated via
appropriate code instrumentation mechanisms to provide a unified system that
offers fault-tolerance capabilities.
The driving principle we will follow is that of minimising the
entanglement between the respective models of the application logic and those of
the recovery logic.
This principled approach to fault-tolerance with clearly delineated
separation of concerns should also manifest itself at the code level of the systems produced,
 that will, in turn, improve the maintainability of the resulting systems.

\section{Conclusions}\label{conc:sec}
% !TEX root = main.tex

We have proposed an approach for the structured development of recovery strategies for distributed applications, realising the choreographed monitor arrangements proposed in earlier work such as \cite{DBLP:conf/pdp/CoppoDV14,FGP12DistribRV,LaneseMZ15}. The proposal build on existing work dealing with runtime adaptations for asynchronous systems on the one hand, and on choreographies projected from global specifications on the other.
An initial step toward the proposed approach, has been  explored  in~\cite{corr/MezzinaT17} where the recovery strategy is directly integrated into the global graph model by specifying some roll-back conditions on choice branches. In this way, during a coordinated choice if the condition does not any more then all the parties involved into the choice are reverted to a state previous the choice, and another  branch can be taken. Starting from the global specification where branches of a choice can be enriched with guards (triggering the possible roll-back)
\textit{controlled reversibility}~\cite{LaneseMSS11} is enforced into local views. Local views are represented by \textit{reversible} CFSMs keeping track of the entire computation history, and local strategies are just a set of constraints to be met in order to revert a certain choice. The framework of~\cite{corr/MezzinaT17} has to be extended in order to allow compliance with monitors implementing a coordinated (and distributed) roll-back. 
Moreover, it could be used as test-bed for different recovery strategies by exploiting the underlying reversible substrate.  Naturally the framework has to be refined in a way to allow for recovery strategies well-formedness
and decidability.

%Moreover, just one global monitor is produced taking care of rolling back the parties involved into a particular branch choice is its condition does not hold anymore. This goes in contrast with the idea of generating a hiararchy of monitors per component, each one implementing a particular strategy interacting with the corresponding monitors of other components.

\nocite{*}
\bibliographystyle{eptcs}
\bibliography{bib}

\begin{thebibliography}{10}
\providecommand{\bibitemdeclare}[2]{}
\providecommand{\surnamestart}{}
\providecommand{\surnameend}{}
\providecommand{\urlprefix}{Available at }
\providecommand{\url}[1]{\texttt{#1}}
\providecommand{\href}[2]{\texttt{#2}}
\providecommand{\urlalt}[2]{\href{#1}{#2}}
\providecommand{\doi}[1]{doi:\urlalt{http://dx.doi.org/#1}{#1}}
\providecommand{\bibinfo}[2]{#2}

\bibitemdeclare{book}{DBLP:books/daglib/0066897}
\bibitem{DBLP:books/daglib/0066897}
\bibinfo{author}{Gul~A. \surnamestart Agha\surnameend} (\bibinfo{year}{1990}):
  \emph{\bibinfo{title}{{ACTORS} - a model of concurrent computation in
  distributed systems}}.
\newblock \bibinfo{series}{{MIT} Press series in artificial intelligence},
  \bibinfo{publisher}{{MIT} Press}, \doi{10.1137/1030027}.

\bibitemdeclare{inproceedings}{DBLP:conf/rv/AttardF16}
\bibitem{DBLP:conf/rv/AttardF16}
\bibinfo{author}{Duncan~Paul \surnamestart Attard\surnameend} \&
  \bibinfo{author}{Adrian \surnamestart Francalanza\surnameend}
  (\bibinfo{year}{2016}): \emph{\bibinfo{title}{A Monitoring Tool for a
  Branching-Time Logic}}.
\newblock In: {\sl \bibinfo{booktitle}{{RV}}}, {\sl \bibinfo{series}{{LNCS}}}
  \bibinfo{volume}{10012}, \bibinfo{publisher}{Springer}, pp.
  \bibinfo{pages}{473--481}, \doi{10.1007/978-3-319-46982-9\_31}.

\bibitemdeclare{article}{bdft16}
\bibitem{bdft16}
\bibinfo{author}{Davide \surnamestart Basile\surnameend},
  \bibinfo{author}{Pierpaolo \surnamestart Degano\surnameend},
  \bibinfo{author}{Gian-Luigi \surnamestart Ferrari\surnameend} \&
  \bibinfo{author}{Emilio \surnamestart Tuosto\surnameend}
  (\bibinfo{year}{2016}): \emph{\bibinfo{title}{Relating two automata-based
  models of orchestration and choreography}}.
\newblock {\sl \bibinfo{journal}{JLAMP}}
  \bibinfo{volume}{85}(\bibinfo{number}{3}), pp. \bibinfo{pages}{425 -- 446},
  \doi{10.1016/j.jlamp.2015.09.011}.

\bibitemdeclare{article}{DBLP:journals/corr/BernardiH15}
\bibitem{DBLP:journals/corr/BernardiH15}
\bibinfo{author}{G.~\surnamestart Bernardi\surnameend} \&
  \bibinfo{author}{M.~\surnamestart Hennessy\surnameend}:
  \emph{\bibinfo{title}{Mutually Testing Processes}}.
\newblock {\sl \bibinfo{journal}{LMCS}} \bibinfo{volume}{11},
  \doi{10.2168/LMCS-11(2:1)2015}.

\bibitemdeclare{article}{bz83}
\bibitem{bz83}
\bibinfo{author}{Daniel \surnamestart Brand\surnameend} \&
  \bibinfo{author}{Pitro \surnamestart Zafiropulo\surnameend}
  (\bibinfo{year}{1983}): \emph{\bibinfo{title}{{On Communicating Finite-State
  Machines}}}.
\newblock {\sl \bibinfo{journal}{Journal of the ACM}}
  \bibinfo{volume}{30}(\bibinfo{number}{2}), pp. \bibinfo{pages}{323--342},
  \doi{10.1145/322374.322380}.

\bibitemdeclare{inproceedings}{DBLP:conf/rv/CassarF15}
\bibitem{DBLP:conf/rv/CassarF15}
\bibinfo{author}{Ian \surnamestart Cassar\surnameend} \&
  \bibinfo{author}{Adrian \surnamestart Francalanza\surnameend}
  (\bibinfo{year}{2015}): \emph{\bibinfo{title}{Runtime Adaptation for Actor
  Systems}}.
\newblock In \bibinfo{editor}{Ezio \surnamestart Bartocci\surnameend} \&
  \bibinfo{editor}{Rupak \surnamestart Majumdar\surnameend}, editors: {\sl
  \bibinfo{booktitle}{RV2015}}, {\sl \bibinfo{series}{LNCS}}
  \bibinfo{volume}{9333}, \bibinfo{publisher}{Springer}, pp.
  \bibinfo{pages}{38--54}, \doi{10.1007/978-3-319-23820-3\_3}.

\bibitemdeclare{inproceedings}{DBLP:conf/ifm/CassarF16}
\bibitem{DBLP:conf/ifm/CassarF16}
\bibinfo{author}{Ian \surnamestart Cassar\surnameend} \&
  \bibinfo{author}{Adrian \surnamestart Francalanza\surnameend}
  (\bibinfo{year}{2016}): \emph{\bibinfo{title}{On Implementing a
  Monitor-Oriented Programming Framework for Actor Systems}}.
\newblock In: {\sl \bibinfo{booktitle}{{IFM} 2016}}, {\sl
  \bibinfo{series}{LNCS}} \bibinfo{volume}{9681},
  \bibinfo{publisher}{Springer}, pp. \bibinfo{pages}{176--192},
  \doi{10.1007/978-3-319-33693-0\_12}.

\bibitemdeclare{article}{DBLP:journals/toplas/CastagnaGP09}
\bibitem{DBLP:journals/toplas/CastagnaGP09}
\bibinfo{author}{G.~\surnamestart Castagna\surnameend},
  \bibinfo{author}{N.~\surnamestart Gesbert\surnameend} \&
  \bibinfo{author}{L.~\surnamestart Padovani\surnameend}
  (\bibinfo{year}{2009}): \emph{\bibinfo{title}{A theory of contracts for Web
  services}}.
\newblock {\sl \bibinfo{journal}{ACM Trans. Program. Lang. Syst.}}
  \bibinfo{volume}{31}(\bibinfo{number}{5}), \doi{10.1145/1538917.1538920}.

\bibitemdeclare{inproceedings}{DBLP:conf/cefp/CesariniT09}
\bibitem{DBLP:conf/cefp/CesariniT09}
\bibinfo{author}{Francesco \surnamestart Cesarini\surnameend} \&
  \bibinfo{author}{Simon~J. \surnamestart Thompson\surnameend}
  (\bibinfo{year}{2009}): \emph{\bibinfo{title}{Erlang Behaviours: Programming
  with Process Design Patterns}}.
\newblock In: {\sl \bibinfo{booktitle}{{CEFP} 2009, Budapest, Hungary}}, pp.
  \bibinfo{pages}{19--41}, \doi{10.1007/978-3-642-17685-2\_2}.

\bibitemdeclare{inproceedings}{chen-jin-meredith-rosu-2009-icicis}
\bibitem{chen-jin-meredith-rosu-2009-icicis}
\bibinfo{author}{Feng \surnamestart Chen\surnameend}, \bibinfo{author}{Dongyun
  \surnamestart Jin\surnameend}, \bibinfo{author}{Patrick \surnamestart
  Meredith\surnameend} \& \bibinfo{author}{Grigore \surnamestart
  Ro\c{s}u\surnameend} (\bibinfo{year}{2009}): \emph{\bibinfo{title}{Monitoring
  Oriented Programming - A Project Overview}}.
\newblock In: {\sl \bibinfo{booktitle}{ICICIS'09}}, \bibinfo{publisher}{ACM},
  pp. \bibinfo{pages}{72--77}.

\bibitemdeclare{inproceedings}{DBLP:conf/pdp/CoppoDV14}
\bibitem{DBLP:conf/pdp/CoppoDV14}
\bibinfo{author}{Mario \surnamestart Coppo\surnameend},
  \bibinfo{author}{Mariangiola \surnamestart Dezani{-}Ciancaglini\surnameend}
  \& \bibinfo{author}{Betti \surnamestart Venneri\surnameend}
  (\bibinfo{year}{2014}): \emph{\bibinfo{title}{Self-Adaptive Monitors for
  Multiparty Sessions}}.
\newblock In: {\sl \bibinfo{booktitle}{22nd Euromicro International Conference
  on Parallel, Distributed, and Network-Based Processing, {PDP} 2014, Torino,
  Italy, February 12-14, 2014}}, \bibinfo{publisher}{{IEEE} Computer Society},
  pp. \bibinfo{pages}{688--696}, \doi{10.1109/PDP.2014.18}.

\bibitemdeclare{inproceedings}{DBLP:conf/fossacs/Francalanza16}
\bibitem{DBLP:conf/fossacs/Francalanza16}
\bibinfo{author}{Adrian \surnamestart Francalanza\surnameend}
  (\bibinfo{year}{2016}): \emph{\bibinfo{title}{{A Theory of Monitors -
  (Extended Abstract)}}}.
\newblock In: {\sl \bibinfo{booktitle}{{FoSSaCS}}}, {\sl
  \bibinfo{series}{LNCS}} \bibinfo{volume}{9634},
  \bibinfo{publisher}{Springer}, pp. \bibinfo{pages}{145--161},
  \doi{10.1007/978-3-662-49630-5\_9}.

\bibitemdeclare{inproceedings}{DBLP:conf/concur/Francalanza17}
\bibitem{DBLP:conf/concur/Francalanza17}
\bibinfo{author}{Adrian \surnamestart Francalanza\surnameend}
  (\bibinfo{year}{2017}): \emph{\bibinfo{title}{{C}onsistently-{D}etecting
  {M}onitors}}.
\newblock In: {\sl \bibinfo{booktitle}{{CONCUR}}}, \bibinfo{series}{LNCS},
  \bibinfo{publisher}{Springer}.
\newblock \bibinfo{note}{(to appear)}.

\bibitemdeclare{article}{FraAI17}
\bibitem{FraAI17}
\bibinfo{author}{Adrian \surnamestart Francalanza\surnameend},
  \bibinfo{author}{Luca \surnamestart Aceto\surnameend} \&
  \bibinfo{author}{Anna \surnamestart Ingolfsdottir\surnameend}
  (\bibinfo{year}{2017}): \emph{\bibinfo{title}{Monitorability for the
  {H}ennessy--{M}ilner logic with recursion}}.
\newblock {\sl \bibinfo{journal}{Formal Methods in System Design}}, pp.
  \bibinfo{pages}{1--30}, \doi{10.1007/s10703-017-0273-z}.

\bibitemdeclare{article}{FGP12DistribRV}
\bibitem{FGP12DistribRV}
\bibinfo{author}{Adrian \surnamestart Francalanza\surnameend},
  \bibinfo{author}{Andrew \surnamestart Gauci\surnameend} \&
  \bibinfo{author}{Gordon~J. \surnamestart Pace\surnameend}
  (\bibinfo{year}{2013}): \emph{\bibinfo{title}{{D}istributed {S}ystem
  {C}ontract {M}onitoring}}.
\newblock {\sl \bibinfo{journal}{The Journal of Logic and Algebraic Programming
  (JLAP)}} \bibinfo{volume}{82}(\bibinfo{number}{5-7}), pp.
  \bibinfo{pages}{186--215}, \doi{10.1016/j.jlap.2013.04.001}.

\bibitemdeclare{article}{FraSey14}
\bibitem{FraSey14}
\bibinfo{author}{Adrian \surnamestart Francalanza\surnameend} \&
  \bibinfo{author}{Aldrin \surnamestart Seychell\surnameend}
  (\bibinfo{year}{2015}): \emph{\bibinfo{title}{Synthesising {C}orrect
  Concurrent {R}untime {M}onitors}}.
\newblock {\sl \bibinfo{journal}{Formal Methods in System Design (FMSD)}}
  \bibinfo{volume}{46}(\bibinfo{number}{3}), pp. \bibinfo{pages}{226--261},
  \doi{10.1007/s10703-014-0217-9}.

\bibitemdeclare{inproceedings}{Gray86}
\bibitem{Gray86}
\bibinfo{author}{Jim \surnamestart Gray\surnameend} (\bibinfo{year}{1986}):
  \emph{\bibinfo{title}{Why Do Computers Stop and What Can Be Done About It?}}
\newblock In: {\sl \bibinfo{booktitle}{SRDS}}, \bibinfo{publisher}{IEEE},
  \doi{10.1109/MC.1983.1654340}.

\bibitemdeclare{inproceedings}{gt16}
\bibitem{gt16}
\bibinfo{author}{Roberto \surnamestart Guanciale\surnameend} \&
  \bibinfo{author}{Emilio \surnamestart Tuosto\surnameend}
  (\bibinfo{year}{2016}): \emph{\bibinfo{title}{An Abstract Semantics of the
  Global View of Choreographies}}.
\newblock In: {\sl \bibinfo{booktitle}{{ICE} 2016, Heraklion, Greece}}, pp.
  \bibinfo{pages}{67--82}, \doi{10.4204/EPTCS.223.5}.

\bibitemdeclare{inproceedings}{RA-SOA:2008}
\bibitem{RA-SOA:2008}
\bibinfo{author}{Florian \surnamestart Irmert\surnameend},
  \bibinfo{author}{Thomas \surnamestart Fischer\surnameend} \&
  \bibinfo{author}{Klaus \surnamestart Meyer-Wegener\surnameend}
  (\bibinfo{year}{2008}): \emph{\bibinfo{title}{Runtime Adaptation in a
  Service-oriented Component Model}}.
\newblock \bibinfo{series}{SEAMS '08}, \bibinfo{publisher}{ACM}, pp.
  \bibinfo{pages}{97--104}, \doi{10.1145/1370018.1370036}.

\bibitemdeclare{phdthesis}{Kalareh:phd}
\bibitem{Kalareh:phd}
\bibinfo{author}{Mehdi~Amoui \surnamestart Kalareh\surnameend}
  (\bibinfo{year}{2012}): \emph{\bibinfo{title}{Evolving Software Systems for
  Self-Adaptation}}.
\newblock Ph.D. thesis, \bibinfo{school}{University of Waterloo},
  \bibinfo{address}{Ontario, Canada}.

\bibitemdeclare{misc}{w3c:cho}
\bibitem{w3c:cho}
\bibinfo{author}{Nickolas \surnamestart Kavantzas\surnameend},
  \bibinfo{author}{Davide \surnamestart Burdett\surnameend},
  \bibinfo{author}{Gregory \surnamestart Ritzinger\surnameend},
  \bibinfo{author}{Tony \surnamestart Fletcher\surnameend} \&
  \bibinfo{author}{Yves \surnamestart Lafon\surnameend} (\bibinfo{year}{2004}):
  \emph{\bibinfo{title}{Web Services Choreography Description Language Version
  1.0}}.
\newblock
  \bibinfo{howpublished}{\url{http://www.w3.org/TR/2004/WD-ws-cdl-10-20041217}}.

\bibitemdeclare{article}{Kell08survey}
\bibitem{Kell08survey}
\bibinfo{author}{Stephen \surnamestart Kell\surnameend} (\bibinfo{year}{2008}):
  \emph{\bibinfo{title}{A Survey of Practical Software Adaptation Techniques.}}
  \bibinfo{volume}{14}(\bibinfo{number}{13}), pp. \bibinfo{pages}{2110--2157}.
\newblock \doi{10.3217/jucs-014-13-2110}.

\bibitemdeclare{inproceedings}{LaneseMSS11}
\bibitem{LaneseMSS11}
\bibinfo{author}{Ivan \surnamestart Lanese\surnameend},
  \bibinfo{author}{Claudio~Antares \surnamestart Mezzina\surnameend},
  \bibinfo{author}{Alan \surnamestart Schmitt\surnameend} \&
  \bibinfo{author}{Jean{-}Bernard \surnamestart Stefani\surnameend}
  (\bibinfo{year}{2011}): \emph{\bibinfo{title}{Controlling Reversibility in
  Higher-Order Pi}}.
\newblock In: {\sl \bibinfo{booktitle}{{CONCUR}}}, pp.
  \bibinfo{pages}{297--311}, \doi{10.1007/978-3-642-23217-6\_20}.

\bibitemdeclare{inproceedings}{lty15}
\bibitem{lty15}
\bibinfo{author}{Julien \surnamestart Lange\surnameend},
  \bibinfo{author}{Emilio \surnamestart Tuosto\surnameend} \&
  \bibinfo{author}{Nobuko \surnamestart Yoshida\surnameend}
  (\bibinfo{year}{2015}): \emph{\bibinfo{title}{{From Communicating Machines to
  Graphical Choreographies}}}.
\newblock In: {\sl \bibinfo{booktitle}{POPL15}}, pp. \bibinfo{pages}{221--232},
  \doi{10.1145/2676726.2676964}.

\bibitemdeclare{inbook}{lty17}
\bibitem{lty17}
\bibinfo{author}{Julien \surnamestart Lange\surnameend},
  \bibinfo{author}{Emilio \surnamestart Tuosto\surnameend} \&
  \bibinfo{author}{Nobuko \surnamestart Yoshida\surnameend}
  (\bibinfo{year}{2017}): \emph{\bibinfo{title}{A tool for choreography-based
  analysis of message-passing software}}.
\newblock \bibinfo{publisher}{{ACM}}.
\newblock \bibinfo{note}{To appear. Available at
  \url{http://www.cs.le.ac.uk/~et52/chorgram_betty_ch.pdf}}.

\bibitemdeclare{article}{meredith-jin-griffith-chen-rosu-2011-jsttt}
\bibitem{meredith-jin-griffith-chen-rosu-2011-jsttt}
\bibinfo{author}{Patrick~O'Neil \surnamestart Meredith\surnameend},
  \bibinfo{author}{Dongyun \surnamestart Jin\surnameend},
  \bibinfo{author}{Dennis \surnamestart Griffith\surnameend},
  \bibinfo{author}{Feng \surnamestart Chen\surnameend} \&
  \bibinfo{author}{Grigore \surnamestart Ro\c{s}u\surnameend}
  (\bibinfo{year}{2011}): \emph{\bibinfo{title}{An Overview of the {MOP}
  Runtime Verification Framework}}.
\newblock {\sl \bibinfo{journal}{International Journal on Software Techniques
  for Technology Transfer}}, pp. \bibinfo{pages}{249--289},
  \doi{10.1007/s10009-011-0198-6}.

\bibitemdeclare{article}{corr/MezzinaT17}
\bibitem{corr/MezzinaT17}
\bibinfo{author}{Claudio~Antares \surnamestart Mezzina\surnameend} \&
  \bibinfo{author}{Emilio \surnamestart Tuosto\surnameend}
  (\bibinfo{year}{2017}): \emph{\bibinfo{title}{Choreographies for Automatic
  Recovery}}.
\newblock {\sl \bibinfo{journal}{CoRR}} \bibinfo{volume}{abs/1705.09525}.

\bibitemdeclare{inproceedings}{ny17}
\bibitem{ny17}
\bibinfo{author}{Rumyana \surnamestart Neykova\surnameend} \&
  \bibinfo{author}{Nobuko \surnamestart Yoshida\surnameend}
  (\bibinfo{year}{2017}): \emph{\bibinfo{title}{Let it recover: multiparty
  protocol-induced recovery}}.
\newblock In: {\sl \bibinfo{booktitle}{CC2017}}, pp. \bibinfo{pages}{98--108},
  \doi{10.1145/3033019.3033031}.

\bibitemdeclare{inproceedings}{Oreizy08SOA}
\bibitem{Oreizy08SOA}
\bibinfo{author}{Peyman \surnamestart Oreizy\surnameend},
  \bibinfo{author}{Nenad \surnamestart Medvidovic\surnameend} \&
  \bibinfo{author}{Richard~N. \surnamestart Taylor\surnameend}
  (\bibinfo{year}{2008}): \emph{\bibinfo{title}{Runtime Software Adaptation:
  Framework, Approaches, and Styles}}.
\newblock \bibinfo{series}{ICSE Companion '08}, \bibinfo{publisher}{ACM},
  \bibinfo{address}{New York, NY, USA}, pp. \bibinfo{pages}{899--910},
  \doi{10.1145/1370175.1370181}.

\bibitemdeclare{book}{roo90}
\bibitem{roo90}
\bibinfo{author}{Paul \surnamestart Rook\surnameend} (\bibinfo{year}{1990}):
  \emph{\bibinfo{title}{Software Reliability Handbook}}.
\newblock \bibinfo{publisher}{Elsevier Science Inc.}, \bibinfo{address}{New
  York, NY, USA}.

\bibitemdeclare{book}{Thomas:2014:PEF:2723830}
\bibitem{Thomas:2014:PEF:2723830}
\bibinfo{author}{Dave \surnamestart Thomas\surnameend} (\bibinfo{year}{2014}):
  \emph{\bibinfo{title}{Programming Elixir: Functional , Concurrent , Pragmatic
  , Fun}}, \bibinfo{edition}{1st} edition.
\newblock \bibinfo{publisher}{Pragmatic Bookshelf}.

\bibitemdeclare{book}{Wyatt:2013:AC:2663429}
\bibitem{Wyatt:2013:AC:2663429}
\bibinfo{author}{Derek \surnamestart Wyatt\surnameend} (\bibinfo{year}{2013}):
  \emph{\bibinfo{title}{Akka Concurrency}}.
\newblock \bibinfo{publisher}{Artima Incorporation}, \bibinfo{address}{USA}.

\end{thebibliography}

\end{document}

\newpage
\listoffixmes

\end{document}

%%% Local Variables: 
%%% mode: latex
%%% TeX-master: t
%%% End: 

%  LocalWords:  choreographies Hypergraphs